\documentclass[final]{IEEEtran}
\pdfoutput=1
\usepackage{amsthm,amssymb,graphicx,multirow,amsmath,color,amsfonts}
\usepackage[update,prepend]{epstopdf}
\usepackage[noadjust]{cite}
\usepackage[latin1]{inputenc}
\usepackage{tikz}
\usepackage{bbm} 
\usepackage{pdfpages}
\usepackage{multirow}
\usepackage{comment}

\def\nbx{{\mathbf{x}}}
\def\nby{{\mathbf{y}}}

\def\nb0{{\mathbf{0}}}
\def\nb1{{\mathbf{1}}}

\def\nbY{{\mathbf{Y}}}

\def\nbbR{{\mathbb{R}}}

\def\nrmc{{\rm c}}
\def\nrmd{{\rm d}}
\def\nrme{{\rm e}}

\def\nrmt{{\rm t}}
\def\nrmu{{\rm u}}

\newtheorem{lemma}{Lemma}
\newtheorem{thm}{Theorem}

\newtheorem{ndef}{Definition}

\newtheorem{prop}{Proposition}
\newtheorem{cor}{Corollary}

\newtheorem{remark}{Remark}

\newtheorem{approximation}{Approximation}	

\def\E{\mathbb{E}}

\def\P{\mathbb{P}}

\def\N{\sigma^2}
\def\T{\beta}							

\def\g{\left.\right|}
\def\bg{\big.\big|}
\def\Bg{\Big.\Big|}

\def\Bgg{\Bigg |}
\def\a{\overset{(a)}{=}}
\def\b{\overset{(b)}{=}}
\def\c{\overset{(c)}{=}}

\allowdisplaybreaks 

\usepackage{setspace}	

\begin{document}
\graphicspath{{./Figures/}}
\title{
Joint Energy and SINR Coverage in Spatially Clustered RF-powered IoT Network
}
\author{
Mohamed A. Abd-Elmagid, Mustafa A. Kishk, and Harpreet S. Dhillon
\thanks{The authors are with Wireless@VT, Department of ECE, Virginia Tech, Blacksburg, VA. Email: \{maelaziz,\ mkishk,\ hdhillon\}@vt.edu. The support of the U.S. NSF (Grants CCF-1464293 and CPS-1739642) is gratefully acknowledged. This paper will be presented in part at the IEEE International Conference on Communications (ICC), 2018 \cite{AbdElmagid2018ICC}. \hfill Manuscript last updated: \today.} 
}

\maketitle

\begin{abstract}
Owing to the ubiquitous availability of radio-frequency (RF) signals, RF energy harvesting is emerging as an appealing solution for powering IoT devices. In this paper, we model and analyze an IoT network which harvests RF energy and receives information from the same wireless network. In order to enable this operation, each time slot is partitioned into {\em charging} and {\em information reception} phases. For this setup, we characterize two performance metrics: (i) energy coverage and (ii) joint signal-to-interference-plus-noise (SINR) and energy coverage. The analysis is performed using a realistic spatial model that captures the spatial coupling between the locations of the IoT devices and the nodes of the wireless network (referred henceforth as the IoT gateways), which is often ignored in the literature. In particular, we model the locations of the IoT devices using a Poisson cluster process (PCP) and assume that some of the clusters have IoT gateways (GWs) deployed at their centers while the other GWs are deployed independently of the IoT devices. The level of coupling can be controlled by tuning the fraction of total GWs that are deployed at the cluster centers. Due to the inherent intractability of computing the distribution of shot noise process for this setup, we propose two accurate approximations, using which the aforementioned metrics are characterized. Multiple system design insights are drawn from our results. For instance, we demonstrate the existence of optimal slot partitioning that maximizes the system throughput. In addition, we explore the effect of the level of coupling between the locations of the IoT devices and the GWs on this optimal slot partitioning. Particularly, our results reveal that the optimal value of time duration for the charging phase increases as the level of coupling decreases. 
\end{abstract}
\begin{IEEEkeywords}
Stochastic geometry, wireless power transmission, Poisson cluster process, coverage probability.
\end{IEEEkeywords}
\section{Introduction} \label{sec:intro}

Due to the massive scale of Internet-of-things (IoT), it is considered highly inefficient and even impractical to replace or recharge batteries of IoT devices especially the ones that are deployed at hard-to-reach places, such as under ground or in tunnels~\cite{7842431,DhiHuaJ2013,DhiHuaJ2014}. This has naturally led to the consideration of energy harvesting to circumvent or supplement conventional power sources, such as replaceable batteries, in these devices. Due to its ubiquity and cost efficient implementation, RF energy harvesting has quickly emerged as an appealing solution for powering IoT devices (majority of which are tiny devices, such as sensors, with very low energy requirement)~\cite{kamalinejad2015wireless}.

The system-level performance analysis of an RF-powered IoT network depends strongly on the choice of the spatial model for the locations of both the RF sources and the IoT devices. Thus far, the existing literature has been mostly limited to the spatial models in which the locations of the IoT devices and RF sources are modeled by two independent point processes, which are usually assumed to be Poisson point processes (PPPs). This is a reasonable first-order choice to study the performance of RF-powered IoT network in which both the IoT devices and the RF sources are deployed fairly uniformly independently of each other in a given region. However, there are several potential IoT deployments in which the IoT devices and RF sources may naturally exhibit strong spatial coupling. One such possibility is when a large number of IoT devices are deployed in certain geographical areas similar to the hotspot zones formed by the humans. In fact, since many IoT applications are related to human assistance (such as health care and smart homes), it is not unreasonable to think that this clustering of IoT devices may be driven by the deployment of more IoT devices in the high population areas. Irrespective of the reason of clustering, it makes sense from the network perspective to deploy RF sources closer to these clusters. We will henceforth refer to these RF sources as IoT GWs, which simply refer to the nodes of the wireless network that is powering the IoT network. For instance, IoT GWs could refer to WiFi access points or small cell base stations. Motivated by this, considering the GWs as the only dedicated source of RF energy in the system, we provide the first analysis for a spatially clustered RF-powered IoT network in which the locations of the IoT devices and the GWs are coupled. Note that as the cluster sizes increases, this setup converges to the independent PPP model used in the literature, which renders the existing results in the literature as special cases of the results derived in this paper. 
\subsection{Related Work}
Energy harvesting wireless networks have been studied in the literature from different perspectives and with the focus on different performance aspects~\cite{UluYenJ2015,Abd-Elmagid2015,Zhang_throughput_maximization,6552840,Abd-Elmagid2016,fullduplex_energycausality,Abd-Elmagid2017,boshkovska2015practical,8292385,7996351}. Recalling the system setup considered in this paper where the locations of the RF-powered IoT devices are assumed to be clustered (in particular, modeled as a PCP), the most relevant literature can be categorized into two sets: (i) stochastic geometry-based analysis of energy harvesting wireless networks and (ii) analysis of wireless networks using PCP. Each of the two categories is discussed next.

{\em Stochastic geometry-based analysis of energy harvesting wireless networks}. Stochastic geometry has been widely used for the analysis of energy harvesting wireless networks due to its tractability and realism~\cite{DhiLiJ2014,7482720,7841754,sakr2015analysis,kishkj1,flint2015performance,conf_version,7876867,huang2014enabling}. The authors in~\cite{DhiLiJ2014} studied the performance of a $K$-tier cellular network in which the BSs are solely powered by energy harvesting. In~\cite{7482720,7841754}, the downlink coverage probability of an RF-powered device was derived.
In both papers, the locations of the users and the BSs were modeled using two independent PPPs. Similar setup was considered in~\cite{sakr2015analysis} with focus on the uplink analysis. In~\cite{kishkj1}, the authors derived a more general performance metric, which is the joint uplink/downlink coverage probability of RF-powered devices.
Although relatively sparse, some works did consider setups in which either the RF-powered users or the BS locations were modeled using a different point process (other than PPP) such as Ginibre $\alpha$-determinantal point process in~\cite{flint2015performance}, Poisson hole process (PHP) in~\cite{conf_version}, and PCP in~\cite{7876867,7974771}. 
The authors in~\cite{7876867} used PCP to model the locations of backscatter transmitters in a backscatter communication system. In particular, they considered a system setup where the backscatter transmitters were clustered around the power beacons.
However, due to the use of beamforming at the power beacons, only the energy received from the cluster center was considered. In~\cite{7974771}, the authors studied the uplink coverage of RF-powered devices with dedicated power beacons. Similar to~\cite{7876867}, only the energy received from the power beacon at the cluster center was considered. In addition, the analysis was focused on SNR coverage instead of SINR, which circumvents some key analytical challenges that result from the high correlation between the interference and the amount of energy harvested from all sources of RF energy. These challenges will be handled carefully in this paper.

{\em Analysis of wireless networks using PCP}. Before going into more details about our contributions, it should be noted that the ability of PCP to capture spatial coupling among different wireless network components has recently made it an appealing choice for modeling the user and/or base station locations in a heterogeneous cellular network (HetNet). In particular, PCP has gained much interest lately for modeling the locations of two types of network components: (i) small cell base stations (SBS) and (ii) mobile users~\cite{7387730,7809177,7110502,7478052,6926852,6810563,7110505}. Both users and SBSs tend to form clusters at the areas of high user density (user hotspots), which makes PCP a more reasonable choice to model their locations. 
For instance, the authors in~\cite{7387730,7809177} used PCP to model the locations of mobile users in cellular networks with the BSs located at the centers of the clusters.
PCP has also been used for modeling the locations of SBSs which are clustered at the locations of high user density to supplement network capacity~\cite{7110502,7478052,6926852,6810563,7110505}. 
Recently, more advanced system setups have been studied where the clustering of both users and SBSs at user hotspots was considered~\cite{mehpcp2017,8023448}.
In~\cite{8023448,chitwc2017}, the authors proposed a unified framework, inspired by 3GPP simulation models, that captures several realistic combinations of spatial distribution of user and SBS locations that appear in real-world HetNet deployments. For this generalized setup, the authors derived the downlink coverage probability for the typical user. Different from these papers, where the focus was on deriving the signal-to-interference-plus-noise ratio (SINR) coverage probability, our paper provides the first analysis of the joint energy and SINR coverage probability for spatially-clustered RF-powered networks.
More details on the contributions in this paper are provided next.
\subsection{Contributions}
This paper studies an RF-powered IoT network, where the IoT GWs are the only source of RF energy. In order to enable this operation, each time-slot is assumed to be divided into two phases: (i) charging phase and (ii) information reception phase. In the charging phase, IoT devices harvest RF energy from the downlink transmissions of the IoT gateways (GWs). In the information reception phase, the IoT devices receive information from the GWs in the downlink channel. For this setup, our main contributions are listed next.

{\em Novel system setup for spatially clustered RF-powered IoT}. Unlike the existing literature where the coupling between the locations of the IoT devices and the GWs is usually ignored, this paper provides a more general setup that captures this coupling. In particular, we assume the locations of the IoT devices to be modeled by clusters where the locations of the cluster centers are modeled using a PPP. To provide a general setup that captures the aforementioned coupling, we assume the locations of the GWs to be modeled using two independent PPPs: (i) the first PPP $\Phi_{\rm b}^{(c)}$ (with density $\lambda_b^{(c)}$) models the locations of the GWs that are deployed at the cluster centers and (ii) the second PPP $\Phi_{\rm b}$ (with density $\lambda_b$) models the locations of the GWs that are randomly deployed in the 2-D plane and are not restricted to lie at the cluster centers. This general setup, as will be shown in the technical part of the paper, captures both the extremes: (i) full coupling between the locations of the IoT devices and the locations of the GWs, which happens when $\lambda_b$ is set to zero and (ii) no coupling between the locations of the IoT devices and the locations of the GWs, which happens when $\lambda_b^{(c)}$ is set to zero. Note that the case of modeling the locations of IoT GWs as an independent PPP, which was commonly used in literature, is a special case of our model, which is equivalent to case (ii). By tuning the values of densities $\lambda_b^{(c)}$ and $\lambda_b$, our model can capture all possible levels of coupling.

{\em Coverage analysis}. This paper provides an accurate characterization of the energy coverage probability of RF-powered IoT when the locations of the IoT devices are modeled as a PCP with a fraction of the total GWs deployed at the cluster centers and the rest deployed as an independent PPP. As will be noted in the technical part, analysis of this setup adds an additional layer of complexity to the derivation of the energy coverage compared to the usual assumption of modeling the locations of IoT devices and the GWs using two independent PPPs. We propose two different approaches to handle this complexity and derive easy-to-use expressions for the energy coverage probability. 
In addition to the energy coverage, we also derive the joint coverage probability, which is the joint probability of harvesting sufficient energy in
the charging phase and achieving strong enough SINR in the information reception phase.
Handling the correlation between both events is also enabled by using the aforementioned two proposed approaches.

{\em System insights}. 
The analysis in this paper provides several useful system design insights. For instance, we show the existence of an optimal duration for the charging phase that maximizes the average system throughput. In addition, we show that this optimal duration of the charging phase increases as the level of coupling between the locations of the IoT devices and the IoT GWs decreases. We also show that deploying IoT GWs at the cluster centers maximizes the joint coverage probability.
\section{System Model}
We study an RF-powered IoT network in which the IoT devices are solely powered by RF energy harvesting circuitries. As discussed in Section~\ref{sec:intro}, inspired by the fact that the GWs are more likely to be deployed in areas where the density of the IoT devices is relatively high, we primarily focus on the setup in which the locations of the IoT devices and GWs are coupled. However, to maintain generality, we consider that the GWs are categorized into two types: i) GWs deployed at the centers of the clusters formed by the IoT devices, and ii) GWs deployed independently from the locations of the IoT devices (e.g., to provide ubiquitous coverage). This generic model enables us to control the level of coupling between the locations of the IoT devices and GWs, by tuning the fraction of total GWs deployed in each type. 
\subsection{Network Model}
 We study a generic scenario in which the locations of IoT devices are modeled by a PCP $\Phi_{\nrmu}$, where the locations of cluster centers are modeled by a PPP $\Phi_{\rm c}$ with density $\lambda_c$. The locations of IoT devices forming each cluster are independent and identically distributed (i.i.d.) given the location of their cluster center~\cite{haenggi2012stochastic}. Union of all locations of IoT devices around cluster centers forms the PCP $\Phi_{\nrmu}$. In particular, there are two types of clusters: i) clusters with GWs deployed at their centers with density $\lambda_{b}^{(c)}$ and average number of IoT devices per cluster $N_1$, and ii) clusters with no GWs deployed at their centers with density $\lambda_c - \lambda_{b}^{(c)}$ and average number of IoT devices per cluster $N_2$. Note that the number of IoT devices in each cluster of $\Phi_{\nrmu}$ is Poisson distributed. It is more probable to have $N_1 > N_2$ owing to the fact that GWs are more likely to be deployed in clusters with a higher number of IoT devices. However, this assumption does not impact the analysis of coverage probabilities, as will be evident in the sequel. The locations of GWs deployed at the cluster centers form a PPP, denoted by $\Phi_{\rm b}^{(c)}$, with density $\lambda_{b}^{(c)}$. We also consider an independent point process of GWs $\Phi_{\rm b}$, which is a PPP with density $\lambda_{b}$. This assumption allows us to tune the level of coupling between the locations of IoT devices and GWs by tuning $\lambda_b^{(c)}$ and/or $\lambda_b$. Thus, the locations of all GWs are simply modeled by a superposition of two independent PPPs and hence form a PPP, denoted by $\Phi$ with density $\lambda$, i.e., $\Phi = \Phi_{\rm b}^{(c)} \cup \Phi_{\rm b}$ and $\lambda = \lambda_{b}^{(c)} + \lambda_{b}$. To maintain generality, the location of an IoT device $u \in \Phi_{\nrmu}$ with respect to its cluster center, denoted by $\nbY_{u} \in \nbbR^2$, is assumed to follow some arbitrary distribution with probability density function $f_{\nbY_{u}}(\cdot)$.

Time is assumed to be slotted with the duration of each slot being $T$ seconds. Each time slot is partitioned into two phases: i) {\em charging phase:} during the first portion of each time slot, $\tau T$ seconds, all GWs act as RF chargers for the IoT devices so that each IoT device could harvest a certain amount of energy required for its communication needs, and ii) {\em information reception phase:} using the harvested energy in the charging phase, each IoT device connects to a certain GW and receives the transmitted data signal by its serving GW during the remaining $\left(1 - \tau\right) T$ seconds.
\subsection{Propagation Model and Metric of Interest}\label{sec2:sub2}
We perform our downlink analysis at a typical IoT device, which is a randomly chosen IoT device from a randomly chosen cluster of $\Phi_{\rm u}$ (referred to as the \textit{representative cluster}, and its center is denoted by $\nbx_{0}$). Due to the stationarity of this setup, the typical IoT device is assumed to be located at the origin without loss of generality. Assuming that the transmitted power by all GWs is the same, denoted by $P_{\nrmt}$, the received power at the location of the typical IoT device from a GW located at $\nbx \in \nbbR^2$ is $P_{\nrmt} g_{\nbx} \lVert \nbx \rVert^{- \alpha}$, where ${g_{\nbx}}$ denotes the small-scale fading gain between the typical IoT device and the GW located at $\nbx$, and $\lVert \nbx \rVert^{- \alpha}$ represents standard power law path-loss with exponent $\alpha > 2$. Under Rayleigh fading assumption, ${g_{\nbx}}$ is an exponential random variable with unit mean, i.e., $g_{\nbx} \sim {\rm exp}(1)$. Hence, the total harvested energy by the typical IoT device from all GWs during charging phase can be expressed as
\begin{align}\label{eq:harvested_energy}
E_{\rm H} = \eta \tau T \sum_{\substack{\nbx \in \Phi}}{P_{\nrmt} g_{\nbx} \lVert \nbx \rVert ^{-\alpha}},
\end{align}
where $0 \leq \eta \leq 1$ is the efficiency of the energy harvesting circuitry \cite{eta_1,eta_2}. The value of $\eta$ depends on the efficiency of the harvesting antenna, the impedance matching circuit and the voltage multipliers. Note that our setup falls in the category of RF-powered wireless networks in which the efficiency of energy harvesting circuitries is assumed to be linear \cite{Abd-Elmagid2016,Zhang_throughput_maximization,Abd-Elmagid2017,fullduplex_energycausality,7996351,7841754}. Incorporating the assumption of having non-linear energy harvesting efficiency~\cite{boshkovska2015practical} in our model is a promising direction for future work. Owing to its longer lifetime compared to regular rechargeable batteries, we assume that a supercapacitor is used for storing the harvested energy at each IoT device. The supercapacitor's large charging and discharging rates make it possible to use the energy soon after it is harvested. However, due to its high leakage current, it is reasonable to assume that any residual energy left in the current time slot may not be available for use in a future time slot~\cite{kim2015energy}. In other words, the energy harvested by each IoT device in a certain time slot is available to be consumed during the same time slot only.

The typical IoT device uses the harvested energy to successfully receive information in the information reception phase under maximum average received power based cell association strategy. In particular, the typical IoT device connects to the GW which provides maximum received power averaged over small-scale fading gain, i.e, it is served by its closest GW. Hence, the signal-to-interference-plus-noise ratio (SINR) at the typical IoT device in the information reception phase can be expressed as 
\begin{align}\label{eq:SINR}
SINR = \frac{P_{\nrmt}h_{\nbx^*}\lVert \nbx^* \rVert^{-\alpha}}{\sigma^2 + \sum_{\substack{\nbx \in \Phi \setminus} \nbx^*}{P_{\nrmt} h_{\nbx} \lVert \nbx \rVert ^{-\alpha}}}, 
\end{align}
where $\nbx^*$ is the location of the serving GW, $h_{\nbx} \sim {\rm exp}(1)$ models the small-scale fading gain in the information reception phase and is assumed to be independent from $g_{\nbx}$, and $\sigma^2$ denotes the thermal noise power. For this setup, we characterize the performance of the RF-powered IoT network in terms of energy coverage probability, joint coverage probability and average downlink achievable throughput, which are formally defined next.
\begin{ndef}\label{def:energy_coverage}
During the charging phase, the energy coverage event occurs when the energy harvested by the typical IoT device is at least $E_{{\rm rec}}$. The typical IoT device needs this amount of energy to power its receiving circuitry and, hence, receive data successfully during the information reception phase. Practically, $E_{{\rm rec}}$ is an increasing function of the target downlink data rate and the duration of the information reception phase~\cite{arafa2015optimal}. The probability of the energy coverage event can be mathematically expressed as
\begin{align}\label{eq:energy_coverage}
E_{\rm cov} = \E \left[ \mathbbm{1} \left(E_{\rm H} \geq E_{\rm rec}\right) \right],
\end{align}
where $\mathbbm{1}(\cdot)$ is the indicator function.
\end{ndef}
\begin{ndef}\label{def:joint_coverage}
The typical IoT device is said to be in joint coverage if two conditions are satisfied: i) $E_{\rm H} \geq E_{\rm rec}$, and ii) the SINR is above a specific threshold value $\beta$ during the information reception phase. Therefore, the joint coverage probability can be mathematically expressed as
\begin{align}\label{eq:joint_coverage}
P_{\rm cov} = \E\left[\mathbbm{1}\left(E_{\rm H} \geq E_{\rm rec}\right) \mathbbm{1}\left(SINR \geq \beta\right)\right].
\end{align}
\end{ndef}
\begin{ndef}\label{def:throughput}
The average received number of bits by the IoT device, per unit time per unit bandwidth, can be expressed as 
\begin{align}
R = \left(1 -\tau\right){\rm log}_2 \left(1 + \beta\right) P_{\rm cov},
\end{align}
where $1 -\tau$ is the fraction of the total time slot duration allocated for information reception phase. 
\end{ndef}
\subsection{Mathematical Preliminaries}
In this subsection, we summarize some key properties of the proposed setup, which will be used throughout the paper. More detailed discussion on these properties can be found in our earlier work~\cite{7809177}, which focused on the downlink SINR coverage of this clustered setup. We will use these properties in this paper to perform the energy and joint coverage analyses. 

In this paper, after deriving the general results in terms of $f_{\nbY_{u}}$ (defined in Section~\ref{sec2:sub2}), we will specialize them to two special PCPs of interest: i) Thomas cluster process, and ii) Mat{\'e}rn cluster process. In a Thomas cluster process~\cite{haenggi2012stochastic}, the devices are distributed according to a normal distribution of variance $\sigma_{\nrmc}$ around their cluster centers ($\Phi_{\rm c}$), which implies
\begin{align}\label{user_distribution_thomas}
f_{\nbY_{u}}(\nby) = \frac{1}{2\pi\sigma_{\nrmc}^2} {\rm exp}\left(-\frac{\lVert \nby \rVert^2}{2 \sigma_{\nrmc}^2}\right),
\end{align}
where $\nby$ is a realization of the random vector $\nbY_{u}$. On the other hand, in a Mat{\'e}rn cluster process~\cite{haenggi2012stochastic}, the locations of devices are sampled uniformly at random independently of each other within a circular disc of radius $R_{\nrmc}$  around their cluster centers, hence
\begin{align}\label{user_distribution_matern}
f_{\nbY_{u}}(\nby) = \begin{cases}
                     \frac{1}{\pi R_{\nrmc}^2},\; &{\rm if}\; \lVert \nby \rVert \leq R_{\nrmc} \\
                     0 &{\rm otherwise}.
                     \end{cases}
\end{align}

For this setup, the overall coverage probability is a combination of the individual coverage probabilities associated with the two potential scenarios: i) the typical IoT device has a deployed GW at its cluster center, and ii) there is no deployed GW at the representative cluster center. For notational simplicity, we describe the performance of the proposed setup in terms of an arbitrary performance metric function. In particular, let $\chi$ $(\bar{\chi})$ denote some arbitrary performance metric function (e.g., energy coverage, joint coverage, or throughput) when there is a deployed GW at the representative cluster center (there is no deployed GW at the representative cluster center), respectively. Now, since each IoT device has an equal chance to be selected as the typical device, the probability that the typical IoT device has a deployed GW at its cluster center is given by
\begin{align}
p_{b} = \frac{N_1 \lambda_{b}^{(c)}}{ N_{1} \lambda_{b}^{(c)} + N_2 \left(\lambda_{c} - \lambda_{b}^{(c)}\right)},
\end{align}
then, from the total probability law, the overall performance can be expressed as 
\begin{align}\label{eq:overall_performance}
\overset{\rm overall}{\chi} = p_{b} \chi + \left(1 - p_b\right) \bar{\chi}.
\end{align}

Note that when there is no GW deployed at the representative cluster center, the location of the typical IoT device becomes independent from the locations of all deployed GWs in the network. Therefore, $\bar{\chi}$ can be mathematically handled in the same way as if the locations of the IoT devices and the GWs are modeled by two independent PPPs, as done in \cite{7841754}. Therefore, in this paper, we will primarily focus on the downlink analysis at a typical IoT device conditioned on the fact that there is a GW located at its cluster center, i.e., our main objective is to derive $\chi$. 

Recall that the locations of the typical IoT device and GW deployed at its cluster center are coupled. In order to explicitly capture this fact, we define two point processes: i) $\Phi_{0}$ which consists of only the representative cluster center, i.e., $\Phi_{0} = \{\nbx_0\}$, and ii) $\Phi_{1}$ which includes the rest of points of $\Phi$, i.e., $\Phi_1 = \Phi \setminus \nbx_0$. By this construction, the link between the typical IoT device and the GW located at its cluster center can be handled separately, as done in~\cite{7809177}. Note that $\Phi_1$ can be argued to have the same distribution as $\Phi$ by applying Slivnyak's theorem~\cite{haenggi2012stochastic}. Since $\Phi_{0}$ includes only the GW located at the representative cluster center, the typical IoT device either connects to the closest GW from $\Phi_{1}$ located at $\nbx_{1}^*$ or the GW located at its cluster center $\nbx_0^* = \nbx_0$. Therefore, the location of the serving GW is given by
\begin{align}
\nbx^* = {\rm arg}~\underset{\nbx \in \{\nbx_0^*,\nbx_1^*\}}{\rm max}~\lVert \nbx \rVert^{-\alpha}.
\end{align}

 \begin{table*}[t!]\caption{Table of notation}
\centering
\begin{center}
\scalebox{.8}{
    \begin{tabular}{ {c} | {c} }
    \hline\hline
    \textbf{Notation} & \textbf{Description} \\ \hline
    $\Phi_{\rm u}$ & PCP modeling the locations of the IoT devices. \\ \hline
    $\Phi_{\rm c}$, $\lambda_{c}$ & PPP modeling the parent point process of $\Phi_{\rm u}$, density of $\Phi_{\rm c}$.\\ \hline
        $\Phi_{\rm b}^{(c)}$, $\lambda_{b}^{(c)}$ & PPP of GWs deployed at a fraction of $\Phi_{\rm c}$ $(\Phi_{\rm b}^{(c)} \subseteq \Phi_{\rm c})$, density of $\Phi_{\rm b}^{(c)}$\\ \hline
        $\Phi$, $\lambda$ & PPP modeling the locations of all GWs deployed in the network, density of $\Phi$\\ \hline
        $N_1$ ($N_2$) & Average number of IoT devices per each cluster with a deployed GW at its center (with no deployed GW at its center).\\ \hline
        $T$, $\tau$, $\eta$ & Duration of each time slot in seconds, fraction of $T$ allocated for charging phase, efficiency of energy harvesting circuitries.\\ \hline
        $P_{\nrmt}$, $\sigma^2$  & Transmit power of all GWs, thermal noise power.\\ \hline
        $g_{\nbx}$, $h_{\nbx}$, $\alpha$ & Rayleigh fading gain in charging phase, Rayleigh fading gain in information reception phase, path-loss exponent.\\ \hline
         $R_i$& Distance between the typical IoT device and its closest GW from $\Phi_i$.\\ \hline
         $W_i$ & Distance from the typical IoT device to its serving GW conditioned on the association with $\Phi_i$.\\ \hline
         $A_i$ &  Probability that the typical IoT device is associated with $\Phi_i$.\\ \hline
          $E_{\rm rec}$, $\T$ & Energy threshold for powering receiving circuitry, SINR threshold for successful demodulation and decoding.\\ \hline
   \hline
    \end{tabular}}
\end{center}
\label{tab:TableOfNotations}
\end{table*}

 Let $R_{i} = \lVert \nbx_{i}^*\rVert$, $i \in \{ 0,1 \}$, denote the distance from the typical IoT device to its closest GW from $\Phi_i$. Then, the distribution of the distance $R_{1}$ is given by~\cite{haenggi2012stochastic}:
\begin{align}\label{R1_pdf}
{\rm PDF}: f_{R_{1}}(r_1) = 2 \pi \lambda~{\rm exp}\left(-\pi \lambda r_1^2\right),~r_{1} \geq 0,
\end{align}
\begin{align}\label{R1_ccdf}
{\rm CCDF}: \bar{F}_{R_1}(r_1) = {\rm exp}\left(- \pi \lambda r_1^2\right),~r_{1} \geq 0.
\end{align}

On the other hand, since the typical IoT device is located at the origin, the relative location of the representative cluster center with respect to the typical IoT device ($\nbx_0$) will have the same distribution as that of the IoT device location $\nbY_{u}$. Therefore, the distribution of $R_0 = \lVert \nbx_{0}^*\rVert$ can be obtained by applying the standard transformation from Cartesian to polar coordinates, to the joint distribution of $\nbx_0$ expressed in Cartesian domain. We provide the distribution of the distance $R_0$ for both Thomas and Mat{\'e}rn cluster processes in the following two remarks~\cite{7809177}.
\begin{remark}\label{R0_distributions_thomas}
If $\Phi_{\rm u}$ is a Thomas cluster process, then the distribution of $R_0$ is given by
\begin{align}\label{R0_pdf_thomas}
{\rm PDF}: f_{R_0}(r_0) = \frac{r_0}{\sigma_{\nrmc}^2} {\rm exp}\left(-\frac{r_0^2}{2 \sigma_{\nrmc}^2}\right),\; r_0 \geq 0,
\end{align}
\begin{align}\label{R0_ccdf_thomas}
{\rm CCDF}: \bar{F}_{R_0}(r_0) = {\rm exp}\left(-\frac{r_0^2}{2\sigma_{\nrmc}^2}\right),\; r_0 \geq 0.
\end{align}
\end{remark}
\begin{remark}\label{R0_distributions_matern}
When $\Phi_{\rm u}$ is a Mat{\'e}rn cluster process, the distribution of $R_0$ is given by
\begin{align}\label{R0_pdf_matern}
{\rm PDF}: f_{R_0}(r_0) = \frac{2r_0}{R_{\nrmc}^2},\;0 \leq r_0 \leq R_{\nrmc},
\end{align}
\begin{align}\label{R0_ccdf_matern}
{\rm CCDF}: \bar{F}_{R_0}(r_0) = \frac{R_{\nrmc}^2 - r_0^2}{R_{\nrmc}^2},\;0 \leq r_0 \leq R_{\nrmc}.
\end{align}
\end{remark}

Let us call $\mathbbm{1}\left({\rm index} = i\right)$ as the association event of the typical IoT device with $\Phi_i$. Given that the typical IoT device is associated with $\Phi_{i}$, the serving distance $W_i$ is the distance between the typical IoT device and its closest GW in $\Phi_i$, i.e., $W_i = R_i \g \mathbbm{1}\left({\rm index} = i\right) = 1$\footnote{Note that this slightly unconventional notation for the conditional random variable is used for notational convenience in the technical exposition. One could of course proceed without this notation by simply absorbing the condition $\mathbbm{1}\left({\rm index} = i\right)$ in the probabilities and expectations.}. Then, the distributions of the serving distance conditioned on the association with $\Phi_{0}$ and $\Phi_{1}$ are given respectively by~\cite{7809177}:
\begin{align}\label{W0_pdf}
{\rm PDF}: f_{W_0}(w_0) = \frac{\bar{F}_{R_1}(w_0) f_{R_0}(w_0)}{A_0},
\end{align}
\begin{align}\label{W1_pdf}
{\rm PDF}: f_{W_1}(w_1) = \frac{\bar{F}_{R_0}(w_1) f_{R_1}(w_1)}{A_1},
\end{align}
where $A_0$ and $A_1$ denote the association probabilities of the typical IoT device with $\Phi_{0}$ and $\Phi_{1}$, respectively, i.e., $A_i = \E\left[\mathbbm{1}\left({\rm index} = i \right)\right]$. The notation used in this paper is summarized in Table~\ref{tab:TableOfNotations}.
\section{Energy Coverage Probability}\label{sec:energy_coverage}
This section is dedicated to studying the energy coverage probability, as defined in Definition \ref{def:energy_coverage}. Deriving an exact closed-form expression for the energy coverage probability is challenging because of the fact that the CDF of the power-law shot noise process, which represents the total amount of harvested energy by the typical IoT device, is not known in closed form~\cite{haenggi2012stochastic}. To lend tractability, we propose two different approximations for this sum, and derive the energy coverage probability associated with each approximation conditioned on the fact that there is a deployed GW at the representative cluster center. Further, we demonstrate that there exists a trade-off between the tightness of the results obtained using those approximations and their tractability. Finally, we characterize the overall energy coverage probability.

Since the typical IoT device is associated with either $\Phi_0$ or $\Phi_1$, from the total probability law, the energy coverage probability, given by (\ref{eq:energy_coverage}), can be expressed as
\begin{align}\label{energy_coverage_detailed}
E_{\rm cov} &= \E\left[\mathbbm{1}\left(E_{\rm H} \geq E_{\rm rec}\right)\right]\nonumber\\ &= \sum_{i=0}^{1}{\E\left[\mathbbm{1}\left(E_{\rm H} \geq E_{\rm rec}\right)\g {\rm index} = i\right] A_{i}}=\sum_{i=0}^{1}{E_{\rm cov}^{(i)} A_{i}}.
\end{align}

\begin{approximation}
The total amount of energy harvested by the typical IoT device is approximated by the energy harvested from the serving GW located at $\nbx^*$ plus the conditional mean of the energy harvested from other GWs. Thus, $E_{\rm H}$ is given by
\begin{align}
E_{\rm H}^{(1)} = \eta \tau T P_{\rm t}\left(g_{\nbx^*} \lVert \nbx^* \rVert^{-\alpha} + \E\left[\sum_{\substack{\nbx \in \Phi \setminus \nbx^* }}{ g_{\nbx} \lVert \nbx \rVert ^{-\alpha}} \Bgg \lVert \nbx^* \rVert \right]\right).
\end{align}
\end{approximation}
\begin{approximation}
The total amount of energy harvested at the typical IoT device is approximated by the energy harvested from the two GWs located at $\nbx_0^*$ and $\nbx_1^*$ plus the conditional mean of the energy harvested from the rest of GWs. Therefore, $E_{\rm H}$ can be expressed as
\begin{align}
&E_{\rm H}^{(2)} = \eta \tau T P_{\rm t}\Bigg( g_{\nbx_1^*} \lVert \nbx_1^* \rVert^{-\alpha} +g_{\nbx_0^*} \lVert \nbx_0^* \rVert^{-\alpha}\nonumber\\ & + \E\left[\sum_{\substack{\nbx \in \Phi \setminus \nbx_1^*, \nbx_0^* }}{ g_{\nbx} \lVert \nbx \rVert ^{-\alpha}} \Bgg \lVert \nbx_1^* \rVert, \lVert \nbx_0^* \rVert \right]\Bigg).
\end{align}
\end{approximation}
In the next two subsections, we derive the energy coverage probability under each approximation.
\subsection{Energy Coverage Probability under Approximation 1}
Under Approximation 1, the energy coverage probability conditioned on the association of the typical IoT device with $\Phi_i$, $i \in \{0,1\}$, is given by the following two Lemmas.
\begin{lemma}\label{lem:E_cov_tier1_approx1}
Given that the typical IoT device associates with $\Phi_1$, the energy coverage probability conditioned on $\Phi$ and under Approximation 1 is given by
\begin{align}\label{eq:E_cov_tier1_approx1_phi}
E_{{\rm cov} \g \Phi}^{(1)} &= \P\left(E_{\rm H} \geq E_{\rm rec}\g {\rm index}=1, \Phi\right) \nonumber\\ &\overset{(1)}{\approx} {\rm e}^{-\left[w_{1}^{\alpha} \left(C(\tau) - \Psi(w_1)\right)\right]^{+}},
\end{align}
while the unconditional probability is given by
\begin{align}\label{eq:E_cov_tier1_approx1}
E_{{\rm cov}}^{(1)} &= \P\left(E_{\rm H} \geq E_{\rm rec}\g {\rm index}=1\right) \nonumber\\ & \overset{(1)}{\approx} \int_{0}^{\infty}{{\rm e}^{-\left[w_{1}^{\alpha} \left(C(\tau) - \Psi(w_1)\right)\right]^{+}}} f_{W_1}(w_1)\nrmd w_1,
\end{align}
where $C(\tau) = \frac{E_{\rm rec}}{\eta \tau T P_{\rm t}}$, $[x]^+ = {\rm max}\{0,x\}$ and $\Psi(w_1)$ is defined as 
\begin{align}\label{eq:epsi_general}
\Psi(w_1)= \int_{r_0 > w_1}^{\infty} r_{0}^{-\alpha}\frac{f_{R_{0}}(r_0)}{\bar{F}_{R_{0}}(w_1)} \nrmd r_{0} + \frac{2 \pi \lambda}{\alpha - 2} w_1^{2 - \alpha}.
\end{align}
\end{lemma}

\begin{IEEEproof}
See Appendix~\ref{app:E_cov_tier1_approx1}.
\end{IEEEproof}

\begin{lemma}\label{lem:E_cov_tier0_approx1}
Given that the typical IoT device associates with $\Phi_0$, the energy coverage probability conditioned on $\Phi$ and under Approximation 1 is given by
\begin{align}\label{eq:E_cov_tier0_approx1_Phi}
E_{{\rm cov} \g \Phi}^{(0)} &= \P\left(E_{\rm H} \geq E_{\rm rec}\g {\rm index}=0, \Phi\right) \nonumber\\ &\overset{(1)}{\approx} {\rm e}^{-\left[w_{0}^{\alpha} \left(C(\tau) - \theta(w_0)\right)\right]^{+}},
\end{align}
while the unconditional probability is given by
\begin{align}\label{eq:E_cov_tier0_approx1}
E_{{\rm cov}}^{(0)} &= \P\left(E_{\rm H} \geq E_{\rm rec}\g {\rm index}=0\right) \nonumber\\ & \overset{(1)}{\approx} F_{W_0}(A) + \int_{A}^{\infty}{{\rm e}^{-\left(C(\tau) w_{0}^{\alpha} - \frac{2 \pi \lambda w_0^2}{\alpha - 2} \right)}} f_{W_0}(w_0) \nrmd w_0,
\end{align}
where $A = \left(\frac{2 \pi \lambda}{C(\tau) \left(\alpha - 2\right)}\right)^{\frac{1}{\alpha - 2}}$ and $\theta(w_0) = \frac{2 \pi \lambda}{\alpha - 2} w_0^{2 - \alpha}$.
\end{lemma}

\begin{IEEEproof}
See Appendix~\ref{app:E_cov_tier0_approx1}.
\end{IEEEproof}
\begin{remark}\label{rem:tau_energycoverage}
Intuitively, increasing the allocated portion of time slot for charging phase, i.e., $\tau T$, allows the IoT devices to harvest more energy during the charging phase and, hence, the energy coverage probability increases. This can be clearly seen from (\ref{eq:E_cov_tier1_approx1}) and (\ref{eq:E_cov_tier0_approx1}), where as $\tau$ increases, $C(\tau)$ decreases and, hence, the energy coverage probability increases. 
\end{remark}
From the results given by Lemmas~\ref{lem:E_cov_tier1_approx1} and \ref{lem:E_cov_tier0_approx1}, the unconditional energy coverage probabilities for Thomas and Mat{\'e}rn cluster processes are presented in the next two corollaries.
\begin{cor}\label{cor:E_cov_approx1_Thomas}
When $\Phi_{\rm u}$ is a Thomas cluster process, the unconditional energy coverage probabilities under Approximation 1 are given by
\begin{align}\label{cor:E_cov_tier1_approx1_Thomas}
E_{{\rm cov}}^{(1)} \overset{(1)}{\approx} \frac{1}{A_1} \int_{0}^{\infty}&{{\rm e}^{-\left(\left[w_{1}^{\alpha} \left(C(\tau) - \Psi(w_1)\right)\right]^{+} + \left(\pi \lambda + \frac{1}{2 \sigma_{\nrmc}^2}\right)w_1^2 \right)}} \times\nonumber\\ &2 \pi \lambda w_1\nrmd w_1,
\end{align}
\begin{align}
E_{{\rm cov}}^{(0)} &\overset{(1)}{\approx} \frac{1 - \nrme^{-A^2\left(\pi \lambda + \frac{1}{2\sigma_{\nrmc}^2}\right)}}{A_0\left(1 + 2 \pi \lambda\sigma_{\nrmc}^2\right)}\nonumber\\ &+\frac{1}{A_0} \int_{A}^{\infty}{{\rm e}^{-\left(C(\tau)w_{0}^{\alpha} + \left(\pi \lambda + \frac{1}{2 \sigma_{\nrmc}^2}-\frac{2\pi\lambda}{\alpha-2}\right)w_0^2 \right)}} \frac{w_0}{\sigma_{\nrmc}^2}\nrmd w_0,
\end{align}
where $A_1 = \frac{2 \pi \lambda \sigma_{\nrmc}^2}{1 + 2 \pi \lambda \sigma_{\nrmc}^2}$, $A_0 = 1 - A_1$ and $\Psi(w_1)$ is given by
\begin{align}\label{eq:epsi_Thomas}
\Psi(w_1)= \frac{{\rm exp}\left(\frac{w_1^2}{2 \sigma_c^2}\right)}{\sigma_c^{\alpha} 2^{\frac{\alpha}{2}}} \Gamma\left(1-\frac{\alpha}{2},\frac{w_1^2}{2\sigma_c^2}\right) + \frac{2 \pi \lambda}{\alpha - 2} w_1^{2 - \alpha}.
\end{align}
\end{cor}

\begin{IEEEproof}
For a Thomas Cluster process, the conditional mean of energy harvested by the typical IoT device from all GWs except the serving one can be obtained as follows
\begin{align}\label{E_mean_tier1_thomas}
\Psi(w_1) &= \int_{r_0 > w_1}^{\infty} r_{0}^{-\alpha}\frac{f_{R_{0}}(r_0)}{\bar{F}_{R_{0}}(w_1)} \nrmd r_{0} + \frac{2 \pi \lambda}{\alpha - 2} w_1^{2 - \alpha}\nonumber\\ & \a  \int_{r_0 > w_1}^{\infty} r_{0}^{-\alpha}\frac{\frac{r_0}{\sigma_c^2} {\rm exp}\left(\frac{-r_0^2}{2 \sigma_c^2}\right)}{{\rm exp}\left(\frac{-w_1^2}{2\sigma_c^2}\right)} \nrmd r_{0} + \frac{2 \pi \lambda}{\alpha - 2} w_1^{2 - \alpha} \nonumber\\
 &\b  \frac{{\rm exp}\left(\frac{w_1^2}{2\sigma_c^2}\right)}{\sigma_c^{\alpha} 2^{\frac{\alpha}{2}}}\int_{\frac{w_1^2}{2\sigma_c^2}}^{\infty} {z^{\frac{-\alpha}{2}} {\rm exp}(-z)} \nrmd z + \frac{2 \pi \lambda}{\alpha - 2} w_1^{2 - \alpha}\nonumber\\ &= \frac{{\rm exp}\left(\frac{w_1^2}{2 \sigma_c^2}\right)}{\sigma_c^{\alpha} 2^{\frac{\alpha}{2}}} \Gamma\left(1-\frac{\alpha}{2},\frac{w_1^2}{2\sigma_c^2}\right) + \frac{2 \pi \lambda}{\alpha - 2} w_1^{2 - \alpha},
\end{align}
where (a) follows from (\ref{R0_pdf_thomas}) and (\ref{R0_ccdf_thomas}), and (b) follows from the change of variables $z = \frac{r_0^2}{2 \sigma_c^2}$. The final expressions are obtained by substituting conditional serving distance distributions from (\ref{W0_pdf}) and (\ref{W1_pdf}) into (\ref{eq:E_cov_tier0_approx1}) and (\ref{eq:E_cov_tier1_approx1}), respectively, along with taking into account that $F_{W_0}(w_0) = 1 - \nrme^{- w_0^2 \left(\pi \lambda + \frac{1}{2 \sigma_{\nrmc}^2}\right)}$.
\end{IEEEproof}
\begin{cor}\label{cor:E_cov_approx1_Matern}
When $\Phi_{\rm u}$ is a Mat{\'e}rn cluster process, the unconditional energy coverage probabilities under Approximation 1 are given by
\begin{align}\label{eq:E_cov_tier1_matern}
E_{{\rm cov}}^{(1)} \overset{(1)}{\approx} \frac{1}{A_1} \int_{0}^{R_{\nrmc}}&{\rm e}^{-\left(\left[w_{1}^{\alpha} \left(C(\tau) - \Psi(w_1)\right)\right]^{+} + \pi \lambda w_1^2 \right)}\times\nonumber\\ &2 \pi \lambda w_1 \frac{R_{\nrmc}^2 - w_1^2}{R_{\nrmc}^2} \nrmd w_1,
\end{align}
\begin{align}\label{eq:E_cov_tier0_matern}
E_{{\rm cov}}^{(0)} \overset{(1)}{\approx} \begin{cases}
\frac{1 - \nrme^{-\pi \lambda A^2 }}{\pi \lambda R_{\nrmc}^2 A_0}+\frac{1}{A_0} \int_{A}^{R_{\nrmc}}&{\rm e}^{-\left(C(\tau)w_{0}^{\alpha} + \left(\pi \lambda -\frac{2\pi\lambda}{\alpha-2}\right)w_0^2 \right)}\times\\ &\frac{2w_0}{R_{\nrmc}^2} \nrmd w_0,\; R_{\nrmc} \geq A\\
\frac{1 - \nrme^{-\pi \lambda R_{\nrmc}^2 }}{\pi \lambda R_{\nrmc}^2 A_0},\; R_{\nrmc} < A
\end{cases}
\end{align}
where $A_1 = \frac{\nrme^{-\pi\lambda R_{\nrmc}^2} + \pi \lambda R_{\nrmc}^2 - 1}{\pi \lambda R_{\nrmc}^2}$, $A_0 = 1 - A_1$ and $\Psi(w_1)$ is given by
\begin{align}\label{eq:epsi_Matern}
\Psi(w_1) = \frac{2}{\alpha - 2}\left(\frac{w_1^{2-\alpha} - R_{\nrmc}^{2-\alpha}}{R_{\nrmc}^{2} - w_1^2}+ \pi \lambda w_1^{2-\alpha}\right).
\end{align}
Further, for the case of $\alpha = 4$, simpler expressions can be obtained as follows
\begin{align}\label{eq:E_cov_tier1_matern_alpha4}
E_{{\rm cov}}^{(1)} \overset{(1)}{\approx} \begin{cases}
&\frac{1}{A_1}\int_{B}^{R_{\nrmc}}{\nrme^{-\left(C(\tau)w_1^4-\frac{w_1^2}{R_{\nrmc}^2}\right)}2\pi \lambda w_1\frac{R_{\nrmc}^2 - w_1^2}{R_{\nrmc}^2}}\nrmd w_1\\&+\frac{\pi \lambda R_{\nrmc}^2 - 1 +\left(\pi \lambda \left(B^2-R_{\nrmc}^2\right) + 1\right)\nrme^{-\pi \lambda B^2}}{\pi \lambda R_{\nrmc}^2 A_1},\; R_{\nrmc} \geq B\\&
\frac{\pi \lambda R_{\nrmc}^2 - 1 + \nrme^{-\pi \lambda R_{\nrmc}^2}}{\pi \lambda R_{\nrmc}^2 A_1},\; R_{\nrmc} < B
\end{cases}
\end{align}
\begin{align}\label{eq:E_cov_tier0_matern_alpha4}
E_{{\rm cov}}^{(0)} \overset{(1)}{\approx} \begin{cases}
&\frac{1}{\pi \lambda R_{\nrmc}^2 A_0} \Bigg[1 - \nrme^{- \pi \lambda A^2} + \frac{\pi \lambda}{2 \sqrt{C(\tau)}} \Bigg(\Gamma\left(\frac{1}{2},C(\tau)A^4\right)\\&-\Gamma\left(\frac{1}{2},C(\tau)R^4\right)\Bigg)\Bigg],\; R_{\nrmc} \geq A\\&
\frac{1 - \nrme^{-\pi \lambda R_{\nrmc}^2 }}{\pi \lambda R_{\nrmc}^2 A_0},\; R_{\nrmc} < A
\end{cases}
\end{align}
\end{cor}

\begin{IEEEproof}
For a Mat{\'e}rn Cluster process, $\Psi(w_1)$ can be derived as follows
\begin{align}\label{E_mean_tier1_matern}
\Psi(w_1) &= \int_{r_0 > w_1}^{\infty} r_{0}^{-\alpha}\frac{f_{R_{0}}(r_0)}{\bar{F}_{R_{0}}(w_1)} \nrmd r_{0} + \frac{2 \pi \lambda}{\alpha - 2} w_1^{2 - \alpha}\nonumber\\& \a  \int_{r_0 > w_1}^{\infty} r_{0}^{-\alpha}\frac{\frac{2r_0}{R_{\nrmc}^2}}{\frac{R_{\nrmc}^2 - w_1^2}{R_{\nrmc}^2}} \nrmd r_{0} + \frac{2 \pi \lambda}{\alpha - 2} w_1^{2 - \alpha} \nonumber\\
& = \frac{2\left(w_1^{2-\alpha} - R_{\nrmc}^{2-\alpha}\right)}{\left(\alpha - 2\right)\left(R_{\nrmc}^2 - w_1^2\right)} + \frac{2 \pi \lambda}{\alpha - 2} w_1^{2 - \alpha},
\end{align}
where (a) follows from (\ref{R0_pdf_matern}) and (\ref{R0_ccdf_matern}). For the case of $\alpha = 4$, (\ref{E_mean_tier1_matern}) reduces to
\begin{align}\label{E_mean_tier1_matern_alpha4}
\Psi(w_1)= \frac{\pi \lambda + \frac{1}{R_{\nrmc}^2}}{w_1^2}.
\end{align}
Substituting (\ref{E_mean_tier1_matern_alpha4}) into (\ref{eq:E_cov_tier1_approx1_phi}), we obtain the following condition on $E^{(1)}_{{\rm cov} \g \Phi}$
\begin{align}\label{condition_mater_alpha4}
E_{{\rm cov} \g \Phi}^{(1)} = 
          \begin{cases}
           \nrme^{-\left(C(\tau) w_1^{\alpha} - \frac{\pi \lambda + \frac{1}{R_{\nrmc}^2}}{w_1^{2- \alpha}}\right)},\;{\rm if} \; w_1 \geq B \\
            1,\;{\rm if} \; w_1 < B
          \end{cases}
\end{align}
where $B =\left(\frac{\pi \lambda + \frac{1}{R_{\nrmc}^2}}{C(\tau)}\right)^{\frac{1}{2}}$. The final result in (\ref{eq:E_cov_tier1_matern_alpha4}) is obtained by plugging the condition in (\ref{condition_mater_alpha4}) into (\ref{eq:E_cov_tier1_matern}) and the final expression in (\ref{eq:E_cov_tier0_matern_alpha4}) follows from applying the change of variables $z = C(\tau) w_0^4$ to the integral in (\ref{eq:E_cov_tier0_matern}). 
\end{IEEEproof}
\begin{remark}\label{rem:4}
In the case of a Thomas cluster process, it can be noticed that as $\sigma_{\nrmc} \rightarrow \infty$, the association probability of the typical IoT device with $\Phi_1$, denoted by $A_1$, approaches 1 and $\Psi(w_1)$, given by (\ref{eq:epsi_Thomas}), approaches $\frac{2 \pi \lambda w_1^{2 - \alpha}}{\alpha - 2}$. Similarly, for a Mat{\'e}rn cluster process, as $R_{\nrmc} \rightarrow \infty$, $A_1$ approaches 1 and $\Psi(w_1)$, given by (\ref{eq:epsi_Matern}), approaches $\frac{2 \pi \lambda w_1^{2 - \alpha}}{\alpha - 2}$. 
\end{remark}
Using Lemmas~\ref{lem:E_cov_tier1_approx1} and \ref{lem:E_cov_tier0_approx1}, the energy coverage probability under Approximation 1 is formally stated in the following Theorem.
\begin{thm}\label{thm:E_cov_approx1}
The energy coverage probability under Approximation 1 can be obtained as
\begin{align}\label{eq:E_cov_approx1}
E_{\rm cov} \overset{(1)}{\approx} A_{0} E_{\rm cov}^{(0)} + A_1 E_{\rm cov}^{(1)},
\end{align}
where $E_{\rm cov}^{(1)}$ and $E_{\rm cov}^{(0)}$ are given respectively by (\ref{eq:E_cov_tier1_approx1}) and (\ref{eq:E_cov_tier0_approx1}).
\end{thm}
\subsection{Energy Coverage Probability under Approximation 2}
Now, we provide the analysis of obtaining the energy coverage probability under Approximation 2. Considering Approximation 2, the conditional energy coverage probabilities are provided in the next two Lemmas.
\begin{lemma}\label{lem:E_cov_tier1_approx2}
Conditioned on the association of the typical IoT device with $\Phi_1$, the energy coverage probability under Approximation 2 is given by
\begin{align}\label{eq:E_cov_tier1_approx2}
E_{{\rm cov}}^{(1)}& \overset{(2)}{\approx} F_{W_1}(A) + \frac{1}{A_1}\int_{A}^{\infty}\int_{w_1}^{\infty}\Bigg(\frac{w_1^{-\alpha} \nrme^{- w_1^{\alpha} \left(C(\tau) - \Psi(w_1)\right)} }{w_1^{-\alpha} - r_0^{-\alpha}}\nonumber\\&-\frac{ r_0^{-\alpha} \nrme^{- r_0^{\alpha} \left(C(\tau) - \Psi(w_1)\right)}}{w_1^{-\alpha} - r_0^{-\alpha}}\Bigg)f_{R_0}(r_0)f_{R_1}(w_1) \nrmd r_0 \nrmd w_1,
\end{align}
where $\Psi(w_1) = \frac{2 \pi \lambda}{\alpha - 2} w_1^{2 - \alpha}$.
\end{lemma}
\begin{IEEEproof}
See Appendix~\ref{app:E_cov_tier1_approx2}.
\end{IEEEproof}
\begin{lemma}\label{lem:E_cov_tier0_approx2}
Conditioned on the association of the typical IoT device with $\Phi_0$, the energy coverage probability under Approximation 2 is given by
\begin{align}\label{eq:E_cov_tier0_approx2}
&E_{{\rm cov}}^{(0)} \overset{(2)}{\approx} 1 - \frac{\nrme^{- \pi \lambda A^2}}{A_0} + \frac{1}{A_0}\int_{0}^{\infty}\int_{A}^{\infty}\Bigg(\frac{w_0^{-\alpha} \nrme^{- w_0^{\alpha} \left(C(\tau) - \theta (r_1)\right)} }{w_0^{-\alpha} - r_1^{-\alpha}}\nonumber\\&-\frac{r_1^{-\alpha} \nrme^{- r_1^{\alpha} \left(C(\tau) - \theta (r_1)\right)}}{w_0^{-\alpha} - r_1^{-\alpha}}\Bigg)f_{R_1}(r_1)f_{R_0}(w_0) \nrmd r_1 \nrmd w_0,
\end{align}
where $\theta(r_1) = \frac{2 \pi \lambda}{\alpha - 2} r_1^{2 - \alpha}$.
\end{lemma}

\begin{IEEEproof}
The result can be obtained using the same approach used in the proof of Lemma~\ref{lem:E_cov_tier1_approx2}.
\end{IEEEproof}
\begin{remark}
Under Approximation 2, the energy coverage probability $E_{\rm cov}$ is obtained by applying Theorem~\ref{thm:E_cov_approx1} where $E_{\rm cov}^{(1)}$ and $E_{\rm cov}^{(0)}$ are given respectively by (\ref{eq:E_cov_tier1_approx2}) and (\ref{eq:E_cov_tier0_approx2}). Furthermore, the conditional energy coverage probabilities for Thomas and Mat{\'e}rn cluster processes can be obtained by substituting the distributions of $R_{1}$ and $R_{0}$ from (\ref{R1_pdf}), (\ref{R0_pdf_thomas}) and (\ref{R0_pdf_matern}) into (\ref{eq:E_cov_tier1_approx2}) and (\ref{eq:E_cov_tier0_approx2}).
\end{remark}
\begin{remark}\label{rem:6}
By construction, it is expected that the expression for energy coverage probability obtained under Approximation 2 will be relatively tighter than the one obtained under Approximation 1. This is attributed to the fact that, under Approximation 2, the total harvested energy by the typical IoT device is approximated by the energy harvested from the two GWs located at $\nbx_0^*$ and $\nbx_1^*$ plus the conditional mean of the harvested energy from other GWs. On the other hand, under Approximation 1, the total harvested energy at the typical IoT device is only approximated by the harvested energy from the serving GW located at $\nbx^*$ plus the conditional mean of the harvested energy from other GWs. That said, as the cluster size increases, the amount of energy harvested from the GW located at the representative cluster center becomes lower. As a result, the energy coverage probability obtained under Approximation 2 converges to the one obtained under Approximation 1.
\end{remark}
\begin{remark}
By observing the derived energy coverage probability expressions for both Approximations 1 and 2, it is clear that the results obtained under Approximation 2 are relatively more complicated than the ones obtained under Approximation 1. This leads to a trade-off between the tightness of the approximation and its tractability, where the tighter the approximation is, the less tractable its expressions are. However, in the numerical results section, we will demonstrate that both Approximations 1 and 2 are tight enough. Therefore, to maintain tractability, we will proceed by considering Approximation 1 in the rest of our analysis.
\end{remark}
\subsection{Overall Energy Coverage Probability}
In this subsection, we are interested in characterizing the overall energy coverage probability for the generic setup considered in this paper. Using the results obtained in this section along with (\ref{eq:overall_performance}), the overall energy coverage probability is given by the following Theorem.
\begin{thm}\label{thm:overall_E_cov_approx1}
The overall energy coverage probability under Approximation 1 can be expressed as
\begin{align}\label{eq:overall_E_cov_approx1}
\overset{\rm overall}{E_{\rm cov}} = p_{b} E_{\rm cov} + \left(1 - p_b\right) \bar{E}_{\rm cov},
\end{align}
where $E_{\rm cov}$ is given by (\ref{eq:E_cov_approx1}) and $\bar{E}_{\rm cov}$ is given by \cite{7841754}:
\begin{align}\label{eq:energycoverage_PPP}
\bar{E}_{\rm cov} & \overset{(1)}{\approx} 1 - \nrme^{-\pi \lambda \left(\frac{2\pi\lambda}{C(\tau)(\alpha - 2)}\right)^{\frac{2}{\alpha-2}}} \nonumber\\&+ \int_{A}^{\infty}{\nrme^{-\left(C(\tau)r_1^{\alpha} + \left(1 - \frac{2}{\alpha - 2}\right)\pi\lambda r_1^2\right)}2\pi\lambda r_1 \nrmd r_1},
\end{align}
where $C(\tau) = \frac{E_{\rm rec}}{\eta \tau T P_{\rm t}}$ and $A = \left(\frac{2 \pi \lambda}{C(\tau) \left(\alpha - 2\right)}\right)^{\frac{1}{\alpha - 2}}$.
\end{thm}
\begin{remark}\label{rem:energy coverge convergence to PPP}
Based on Remark~\ref{rem:4}, as the cluster size goes to infinity, the energy coverage probability $E_{\rm cov}$, given by Theorem~\ref{thm:E_cov_approx1}, reduces to $\bar{E}_{\rm cov}$. This is due to the fact that when the cluster size goes to infinity, there will be no coupling between the locations of the IoT devices and that of the GWs.    
\end{remark}
\section{Joint Coverage Probability}\label{sec:joint_coverage}
In this section, using the conditional energy coverage probability results obtained in Section~\ref{sec:energy_coverage}, we derive the joint coverage probability given by Definition~\ref{def:joint_coverage}. Afterwards, using the joint coverage probability result, we characterize the average downlink achievable throughput. Finally, we obtain the overall joint coverage probability.

From total probability law, the joint coverage probability can be expressed as
\begin{align}
\nonumber P_{\rm cov} &= \E\left[\mathbbm{1}\left(SINR \geq \T \right)\mathbbm{1}\left(E_{\rm H} \geq E_{\rm rec}\right)\right]\nonumber\\&= \sum_{i=0}^{1}{\E\left[\mathbbm{1}\left(SINR \geq \T \right)\mathbbm{1}\left(E_{\rm H} \geq E_{\rm rec}\right)\g {\rm index} = i\right] A_i}\nonumber\\
&= \sum_{i=0}^{1}{P_{\rm cov}^{(i)} A_i}.
\end{align}
\subsection{Joint Coverage Probability}
In this subsection, our primary objective is to derive the joint coverage probability experienced by the typical IoT device when there is a GW deployed at its cluster center. Towards this objective, we start by deriving the joint coverage probability conditioned on the association of the typical IoT device with $\Phi_i$. Afterwards, we derive the joint coverage probability using the total probability law. The conditional joint coverage probabilities are given by the following two Lemmas.
\begin{lemma}\label{lem:joint_cov_tier1}
Conditioned on the association of the typical IoT device with $\Phi_1$, the joint coverage probability with SINR threshold $\beta$ and energy threshold $E_{\rm rec}$ is given by
\begin{align}\label{eq:joint_cov_tier1}
&P_{\rm cov}^{(1)} = \E_{\Phi}\left[E_{{\rm cov} \bg \Phi}^{(1)} S_{{\rm cov }\g \Phi}^{(1)} \bg {\rm index}=1 \right]
\nonumber\\ &\overset{(1)}{\approx} \frac{1}{A_1}\int_{w_1 = 0}^{\infty} \nrme ^{-\left( \frac{\T \N w_1^{\alpha}}{P_{\nrmt}} + \left[w_{1}^{\alpha} \left(C(\tau) - \Psi(w_1)\right)\right]^{+} + 2 \pi \lambda w_1^2 \rho(\T,\alpha)\right)}\times\nonumber\\& f_{R_1}(w_1)\int_{r_0 > w_1}^{\infty} {\frac{1}{1 + \T w_1^{\alpha} r_0^{-\alpha}}f_{R_{0}}(r_0)}\nrmd r_{0} \nrmd w_1,
\end{align}
where $S_{{\rm cov }\g \Phi}^{(1)}$ is the SINR coverage probability conditioned on $\Phi$ and the association with $\Phi_1$, $\Psi(w_1)$ is given by (\ref{eq:epsi_general}) and $\rho(\T,\alpha)$ is defined as $\rho(\T,\alpha) = \frac{\T^{\frac{2}{\alpha}}}{2} \int_{\T^{\frac{-2}{\alpha}}}^{\infty}{\frac{1}{1 + u^{\frac{\alpha}{2}}} \nrmd u}$.
\end{lemma}

\begin{IEEEproof}
See Appendix~\ref{app:joint_cov_tier1}.
\end{IEEEproof}

\begin{lemma}\label{lem:joint_cov_tier0}
Conditioned on the association of the typical IoT device with $\Phi_0$, the joint coverage probability with SINR threshold $\beta$ and energy threshold $E_{\rm rec}$ is given by
\begin{align}\label{eq:joint_cov_tier0}
&P_{\rm cov}^{(0)} \nonumber = \E_{\Phi}\left[E_{{\rm cov} \bg \Phi}^{(0)} S_{{\rm cov }\g \Phi}^{(0)} \g {\rm index}=0\right]\\
&\overset{(1)}{\approx} \int_{0}^{A}{\nrme^{-\left(\frac{\T \N w_0^{\alpha}}{P_{\nrmt}} + 2 \pi \lambda w_0^2 \rho(\T,\alpha)\right)} f_{W_0}(w_0)}  \nrmd w_0 \nonumber\\&+ \int_{A}^{\infty}{ \nrme^{-\left(\left[\frac{\T \N}{P_{\nrmt}} + C(\tau)\right]w_0^{\alpha} +  \left[\rho(\T,\alpha) - \frac{1}{\alpha - 2}\right] 2 \pi \lambda w_0^2\right)} f_{W_0}(w_0)} \nrmd w_0,
\end{align}
where $S_{{\rm cov }\g \Phi}^{(0)}$ is the SINR coverage probability conditioned on $\Phi$ and the association with $\Phi_0$.
\end{lemma}

\begin{IEEEproof}
See Appendix~\ref{app:joint_cov_tier0}.
\end{IEEEproof}

\begin{remark}\label{rem:tau_jointcoverage}
Similar to Remark~\ref{rem:tau_energycoverage} for the energy coverage probability, from (\ref{eq:joint_cov_tier1}) and (\ref{eq:joint_cov_tier0}), it is clear that increasing $\tau$ decreases $C(\tau)$, and consequently the joint coverage probability increases.
\end{remark}

Using Lemmas~\ref{lem:joint_cov_tier1} and \ref{lem:joint_cov_tier0}, the joint coverage probability is formally stated in the following Theorem.
\begin{thm}\label{thm:joint_cov_approx1}
The joint coverage probability under Approximation 1 can be obtained as
\begin{align}\label{Eq:joint_cov_approx1}
P_{\rm cov} \overset{(1)}{\approx} A_0 P_{\rm cov}^{(0)} + A_1 P_{\rm cov}^{(1)},
\end{align}
where $P_{\rm cov}^{(1)}$ and $P_{\rm cov}^{(0)}$ are given respectively by (\ref{eq:joint_cov_tier1}) and (\ref{eq:joint_cov_tier0}).
\end{thm}
\subsection{Average Downlink Throughput}
Using the joint coverage probability obtained in the previous subsection, the average downlink achievable throughput is characterized in this subsection. Applying Definition \ref{def:throughput}, the average achievable throughput is given by the following proposition.
\begin{prop}\label{prop:1}
The average downlink achievable throughput per IoT device, expressed in bits/sec/Hz, is given by $R = \left(1 -\tau\right) {\rm log}_2\left(1 + \beta\right) P_{\rm cov}$, where $P_{\rm cov}$ is given by Theorem~\ref{thm:joint_cov_approx1}. Here, the fraction $1 - \tau$ is due to the fact that the typical IoT device only receives data from its serving GW during the information reception phase, which occupies $1 - \tau$ fraction of the total time slot duration.
\end{prop}
\begin{remark}\label{rem:9}
As discussed in Remark~\ref{rem:tau_jointcoverage}, $P_{\rm cov}$ is an increasing function of $\tau$. However, the portion of time slot dedicated for information reception phase, i.e., $(1 - \tau) T$, decreases with $\tau$. This suggests the existence of an optimal $\tau$ so as to maximize the average downlink achievable throughput. We will investigate the impact of the cluster size on the optimal slot partitioning policy in the numerical results section.
\end{remark}
\subsection{Overall Joint Coverage Probability}
Now, the overall joint coverage probability can be obtained by the following Theorem.
\begin{thm}\label{thm:overall_joint_cov_approx1}
The overall joint coverage probability under Approximation 1 is given by
\begin{align}\label{eq:overall_joint_cov_approx1}
\overset{\rm overall}{P_{\rm cov}} = p_{b} P_{\rm cov} + \left(1 - p_b\right) \bar{P}_{\rm cov},
\end{align}
where $P_{\rm cov}$ is given by (\ref{Eq:joint_cov_approx1}) and $\bar{P}_{\rm cov}$ is given by \cite{7841754}:
\begin{align}\label{eq:jointcoverage_PPP}
&\bar{P}_{\rm cov} \overset{(1)}{\approx} \int_{0}^{A}{ \nrme ^{-\left( \frac{\T \N r_1^{\alpha}}{P_{\nrmt}} + \pi \lambda r_1^2 + \nu\left(r_1,\T\right)\right)} 2\pi\lambda r_1 \nrmd r_1} + \nonumber\\&\int_{A}^{\infty} {\nrme^{-\left(\left[\frac{\T \N}{P_{\nrmt}} + C(\tau)\right]r_1^{\alpha} +  \left[1 - \frac{2}{\alpha - 2}\right]\pi \lambda r_1^2 + \nu\left(r_1,\T \right)\right)} 2\pi\lambda r_1 \nrmd r_1},
\end{align}
where $\nu\left(r_1,\T \right) = \frac{2 \pi \lambda \beta^{\frac{2}{\alpha}}r_1^2}{\alpha} \int_{\frac{1}{\T}}^{\infty}{\frac{1}{z^{1 - \frac{2}{\alpha}}(1 + z)} \nrmd z}$.
\end{thm}
\begin{remark}\label{rem:joint coverge convergence to PPP}
Similar to Remark~\ref{rem:energy coverge convergence to PPP}, as the cluster size goes to infinity, the joint coverage probability $P_{\rm cov}$, given by Theorem~\ref{thm:joint_cov_approx1}, reduces to $\bar{P}_{\rm cov}$.    
\end{remark}
\section{Discussion and Numerical Results}
Recall that for any performance metric $\chi$ (which can represent the energy coverage, the joint coverage, or the throughput), the value of this performance metric can be calculated using (\ref{eq:overall_performance}), which can be rewritten as follows:
\begin{align}\label{eqn:num1}
\overset{\rm overall}{\chi} = \frac{\zeta \bar{\chi} + \gamma \left(\chi - \zeta \bar{\chi}\right)}{\zeta + \gamma \left(1 - \zeta\right)},
\end{align}
where $\zeta = \frac{N_2}{N_1}$ and $\gamma = \frac{\lambda_{b}^{(c)}}{\lambda_{c}}$. In case $\lambda_{c}=\lambda$, $0 \leq \gamma \leq 1$ is the fraction of clusters with GWs deployed at their centers. Our first objective is to characterize the optimal deployment policy of GWs that maximizes the overall performance of the IoT network $(\overset{\rm overall}{\chi})$. In particular, our target is to find the optimal density of GWs deployed at the cluster centers $\lambda_b^{(c)}$ that maximizes $\overset{\rm overall}{\chi}$. By differentiating the result in (\ref{eqn:num1}) with respect to $\gamma$, the following Remark directly follows.
\begin{remark}\label{rem:15}
The overall performance for the IoT network considered in this paper is maximized when all the GWs are deployed at the cluster centers. This is due to the fact that the derivative of (50) with respect to $\gamma$ is positive since $\chi>\bar{\chi}$. The intuition behind this result can be explained as follows. Consider the two extreme cases: (i) $\lambda_b^{(c)}=0$ and (ii) $\lambda_b^{(c)}=\lambda$. The point process $\Phi_1 = \Phi \setminus \nbx_0$ observed by a typical IoT device is the same in both cases. Hence, the only difference between the two extreme cases is that in (ii) the typical IoT device has a GW located at its cluster center while in (i) it does not. This clearly implies that the performance in (ii) is lower bounded by (i). Obviously, the performance of (ii) will converge to this lower bound as the cluster size increases.
\end{remark}
Now, we verify the accuracy of the expressions derived in Sections~\ref{sec:energy_coverage} and~\ref{sec:joint_coverage} by comparing them with simulation results. We focus on the expressions derived for the performance of the IoT device that has a GW deployed at the center of its cluster, namely, $E_{\rm cov}$, $P_{\rm cov}$, and $R$. This enables us to investigate the system insights resulting from the clustered spatial distribution of the RF-powered IoT devices, which is the main contribution of this paper. Unless otherwise specified, the following simulation setup is considered: $\alpha=4$, $\lambda=0.01$, $E_{\rm rec}= \left(1 - \tau\right) T \left(a R^{\prime} + b\right)$ joules, $a=10^{-4}$, $b=5\times10^{-5}$, $\eta=0.5$, and $P_{\rm t}=1$. 

\begin{figure}[!t]
    \centering
    \begin{minipage}{.41\textwidth}
        \centering
\includegraphics[width=1\columnwidth]{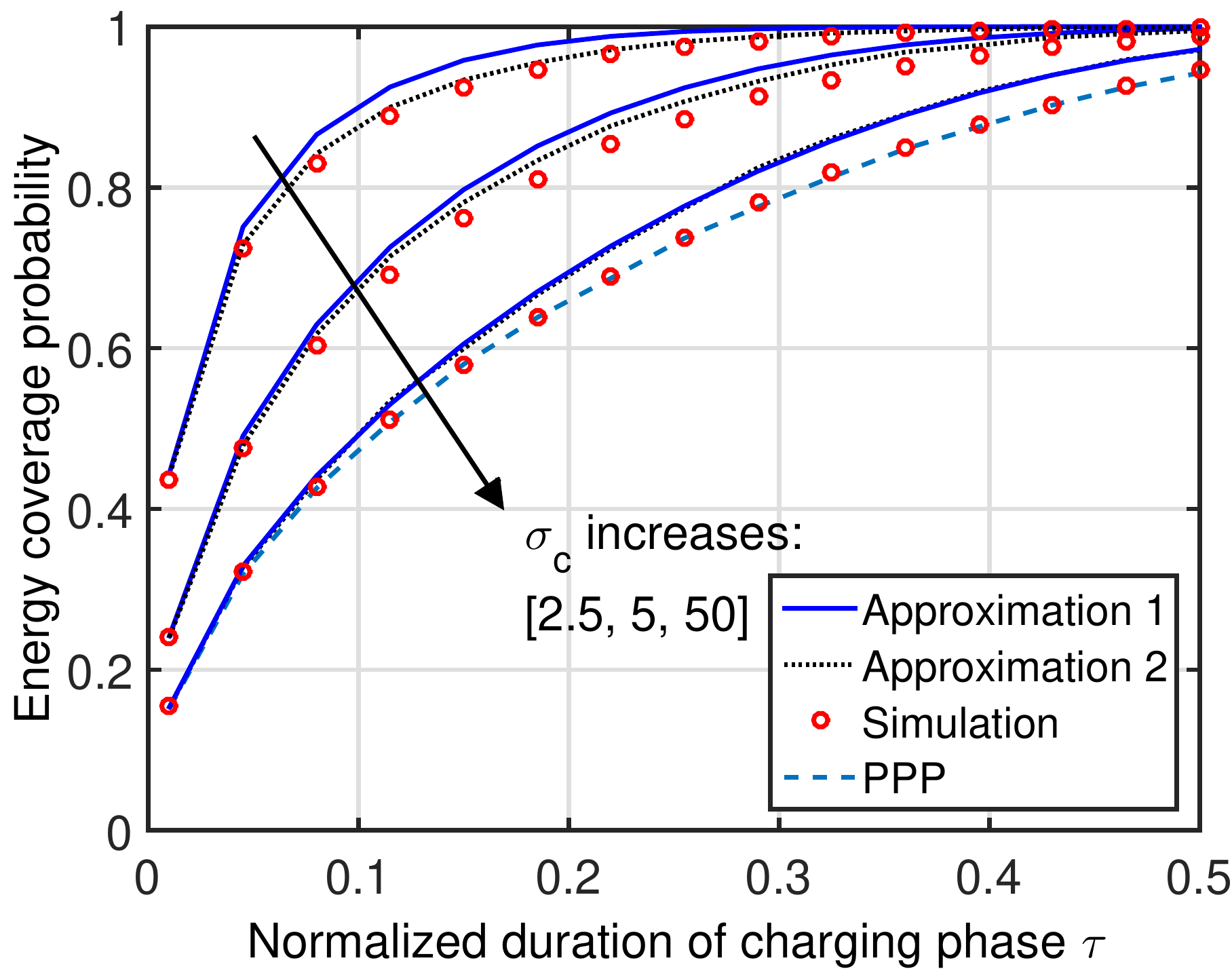}
\caption{The energy coverage probability of a typical IoT device when the representative cluster has a GW deployed at its center for Thomas cluster process.}
\label{fig:1}
    \end{minipage}%
    \hfill
    \begin{minipage}{0.41\textwidth}
        \centering
\includegraphics[width=1\columnwidth]{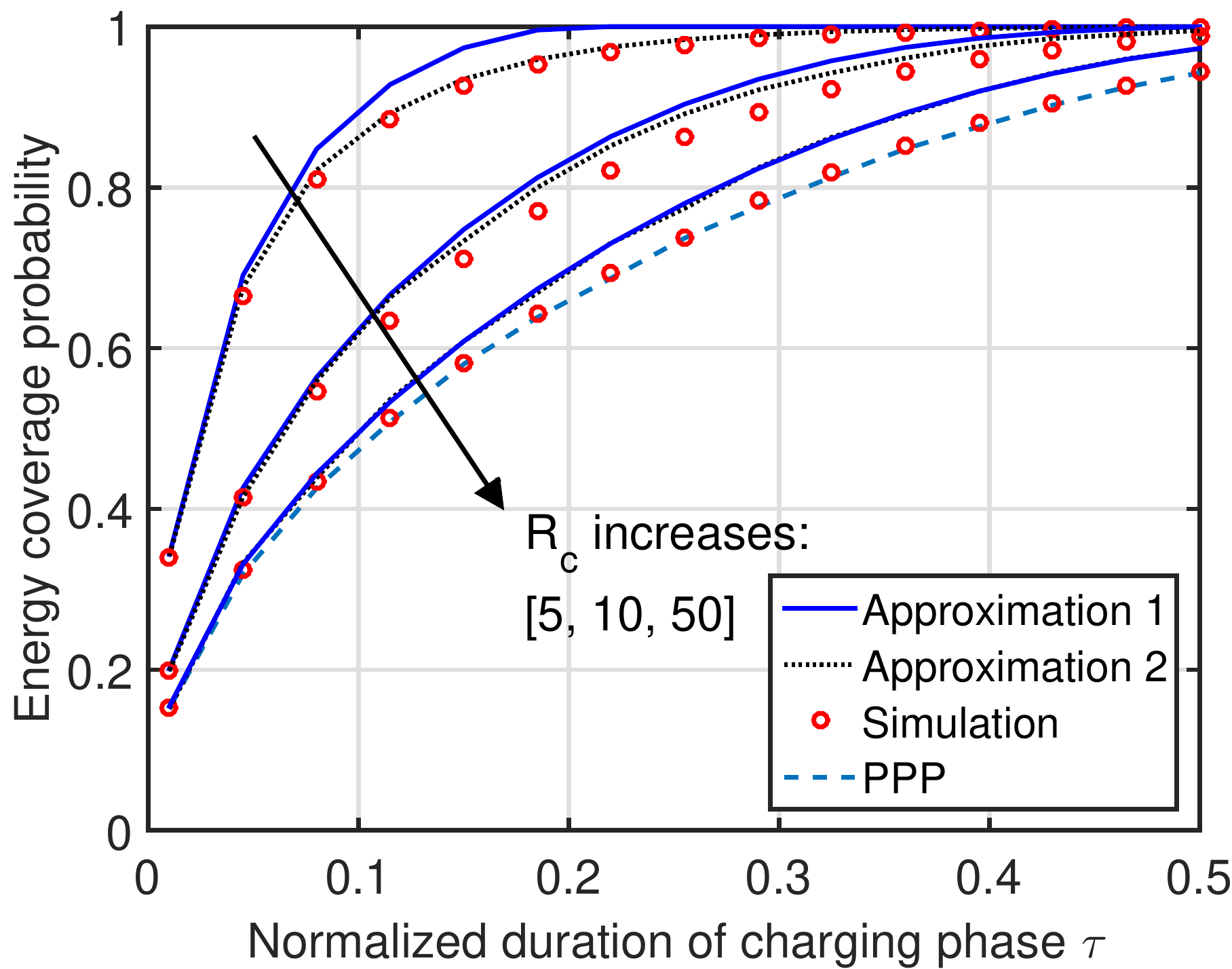}
\caption{The energy coverage probability of a typical IoT device when the representative cluster has a GW deployed at its center for Mat{\'e}rn cluster process.}
\label{fig:2}
    \end{minipage}
\end{figure}

In Figs.~\ref{fig:1} and~\ref{fig:2}, we plot the energy coverage probability for Thomas cluster process and Mat{\'e}rn cluster process, respectively. The results support our comments in Remarks~\ref{rem:tau_energycoverage} and~\ref{rem:tau_jointcoverage} that the energy coverage probability increases as the duration of the charging phase $\tau T$ increases. In addition, as noted in Remark~\ref{rem:6}, Approximation 2 provides relatively tighter results at lower values of $\sigma_c$ ($R_{\rm c}$ in case of Mat{\'e}rn cluster process) compared to Approximation 1. However, as the cluster size increases (which is equivalent to increasing $\sigma_{\rm c}$ or $R_{\rm c}$) we notice that the results from both approximations become almost the same. In addition, we note that the energy coverage probability increases as the values of $\sigma_{\rm c}$ or $R_{\rm c}$ are decreased. Recalling that in our setup the GWs are deployed at the cluster
centers, where the locations of the cluster centers are modeled by a PPP, we compare the performance of our setup with the one in which the locations of the IoT devices and the GWs are modeled using two independent PPPs. The latter setup, which was studied in~\cite{7841754}, is referred to in Fig. 1 as PPP. As expected, the gap between the performance of the considered setup and
the PPP setup from~\cite{7841754} increases as the cluster size decreases. Furthermore, we notice that as the cluster size increases, the energy coverage probability converges to that of the PPP setup of~\cite{7841754}. Note that the impact of the density of gateways on the considered performance metrics will be similar to that of the duration of charging phase $\tau T$. Particularly, as the density of gateways increases, the amount of energy harvested at the typical device increases. Consequently, the energy and joint coverage probabilities will increase as well.

\begin{figure}[!t]
    \centering
    \begin{minipage}{.41\textwidth}
        \centering
\includegraphics[width=1\columnwidth]{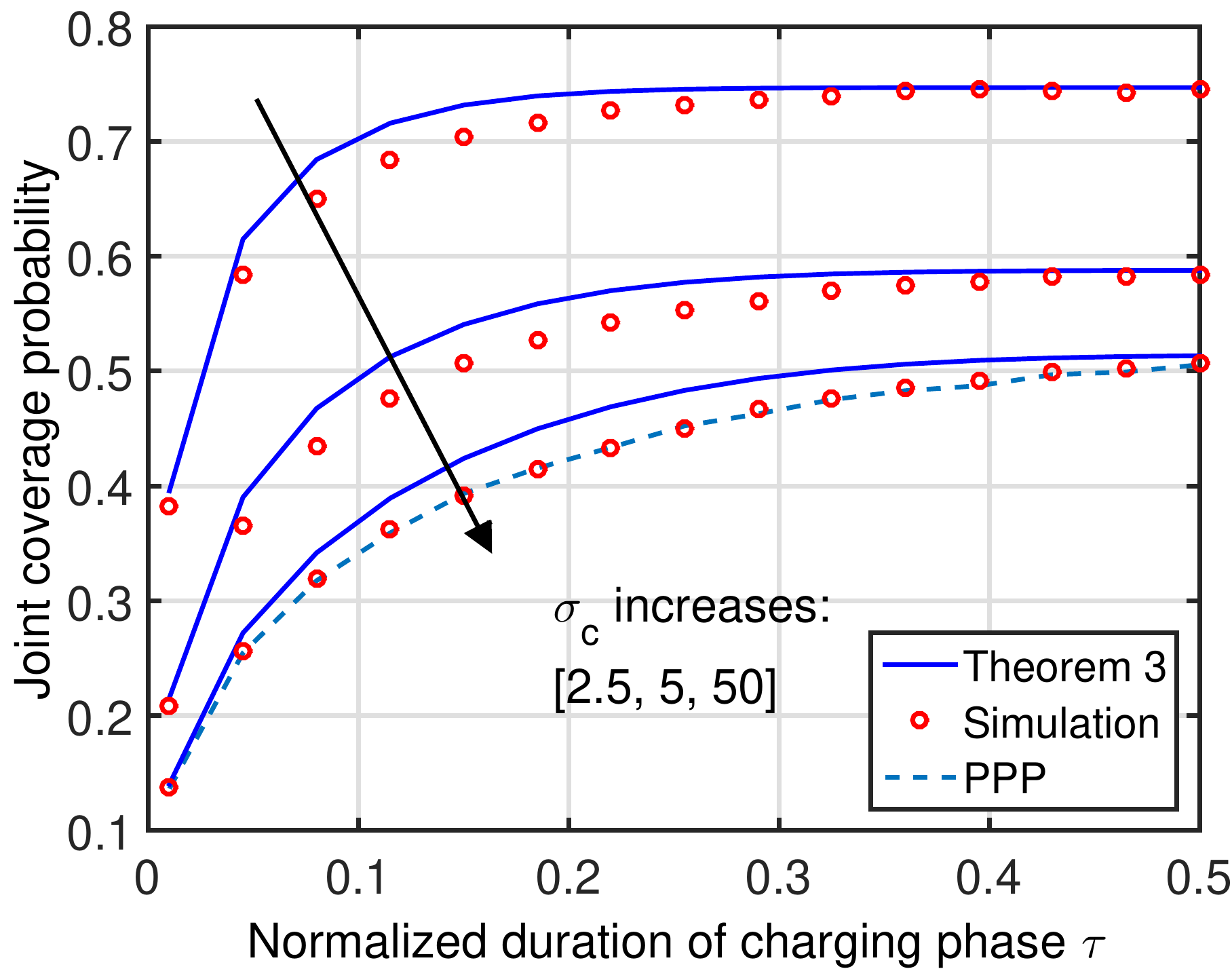}
\caption{The joint coverage probability of a typical IoT device when the representative cluster has a GW deployed at its center for Thomas cluster process.}
\label{fig:3}
    \end{minipage}%
    \hfill
    \begin{minipage}{0.41\textwidth}
        \centering
\includegraphics[width=1\columnwidth]{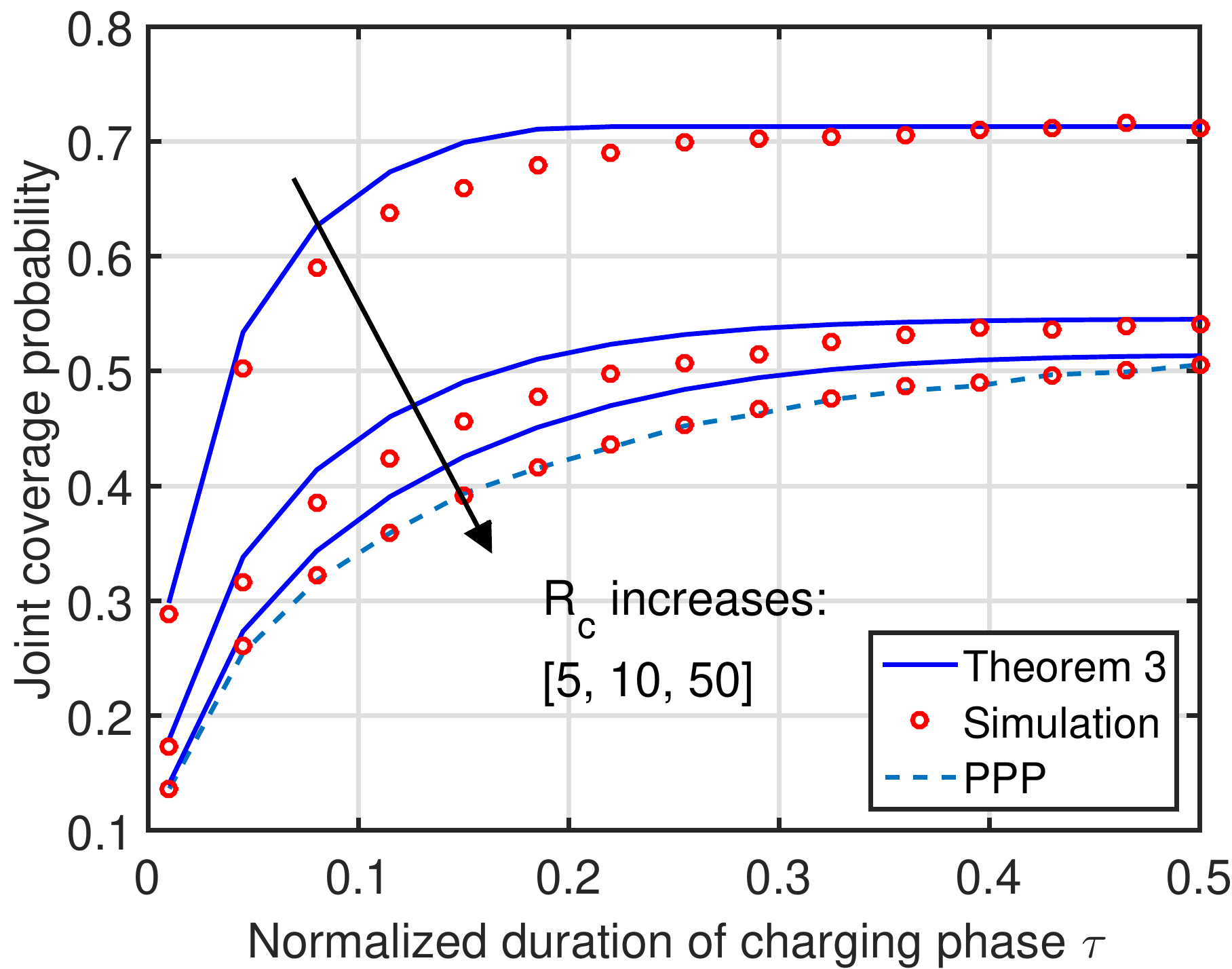}
\caption{The joint coverage probability of a typical IoT device when the representative cluster has a GW deployed at its center for Mat{\'e}rn cluster process.}
\label{fig:4}
    \end{minipage}
\end{figure}

In Figs.~\ref{fig:3} and~\ref{fig:4}, we plot the joint coverage probability derived in Theorem~\ref{thm:joint_cov_approx1} against different values of $\tau$ for Thomas cluster process and Mat{\'e}rn cluster process, respectively. We notice that the joint coverage probability converges to a fixed value as $\tau$ increases. This is expected due to the convergence of the energy coverage probability to unity as $\tau$ increases, which reduces the joint coverage probability to only SINR coverage when $\tau$ is large enough. Similar to the energy coverage, the joint coverage probability converges to the performance of the setup considered~\cite{7841754} as the cluster size increases.

\begin{figure}[!t]
    \centering
    \begin{minipage}{.41\textwidth}
        \centering
\includegraphics[width=1\columnwidth]{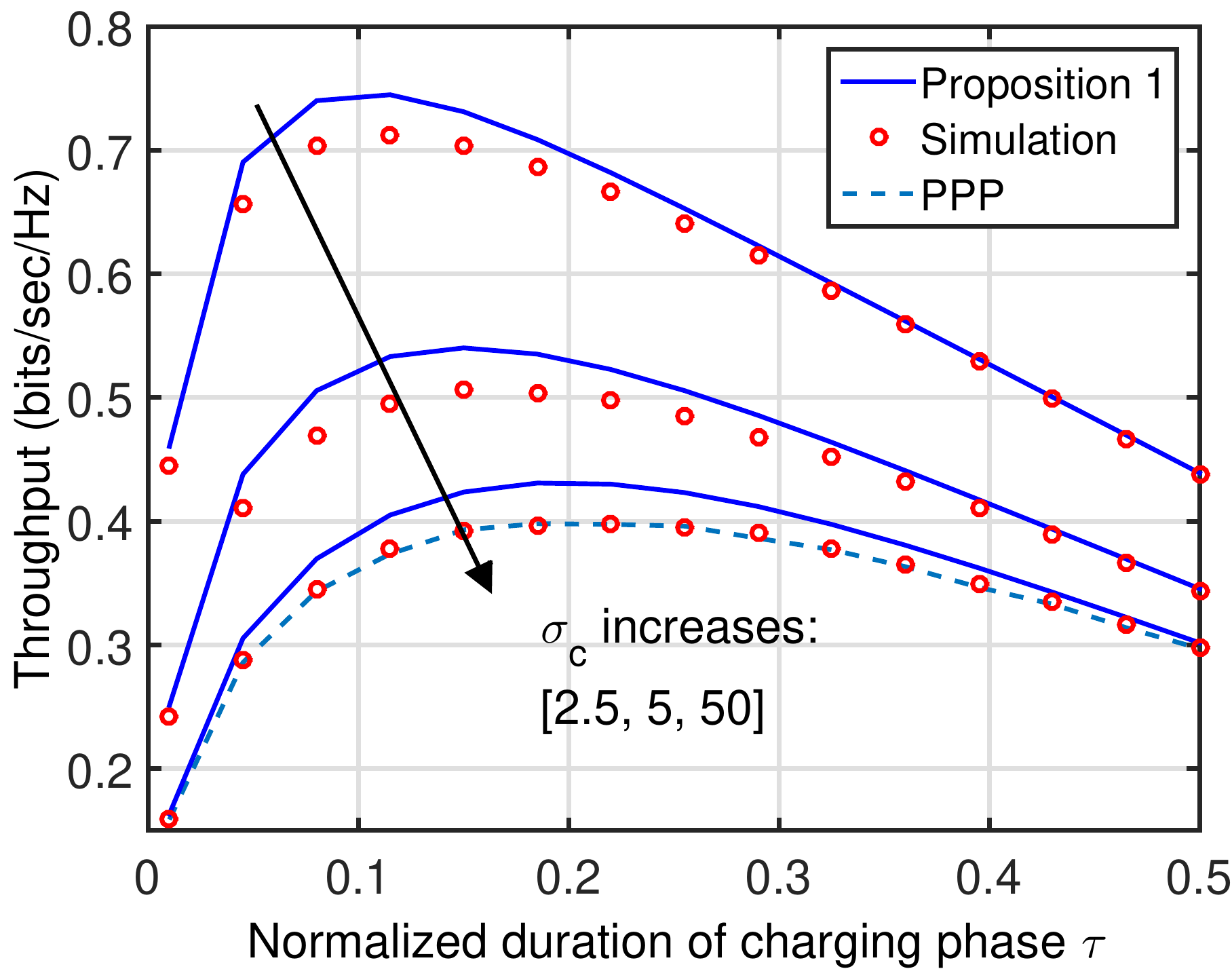}
\caption{The average downlink achievable throughput of a typical IoT device when the representative cluster has a GW deployed at its center for Thomas cluster process.}
\label{fig:5}
    \end{minipage}%
    \hfill
    \begin{minipage}{0.41\textwidth}
        \centering
\includegraphics[width=1\columnwidth]{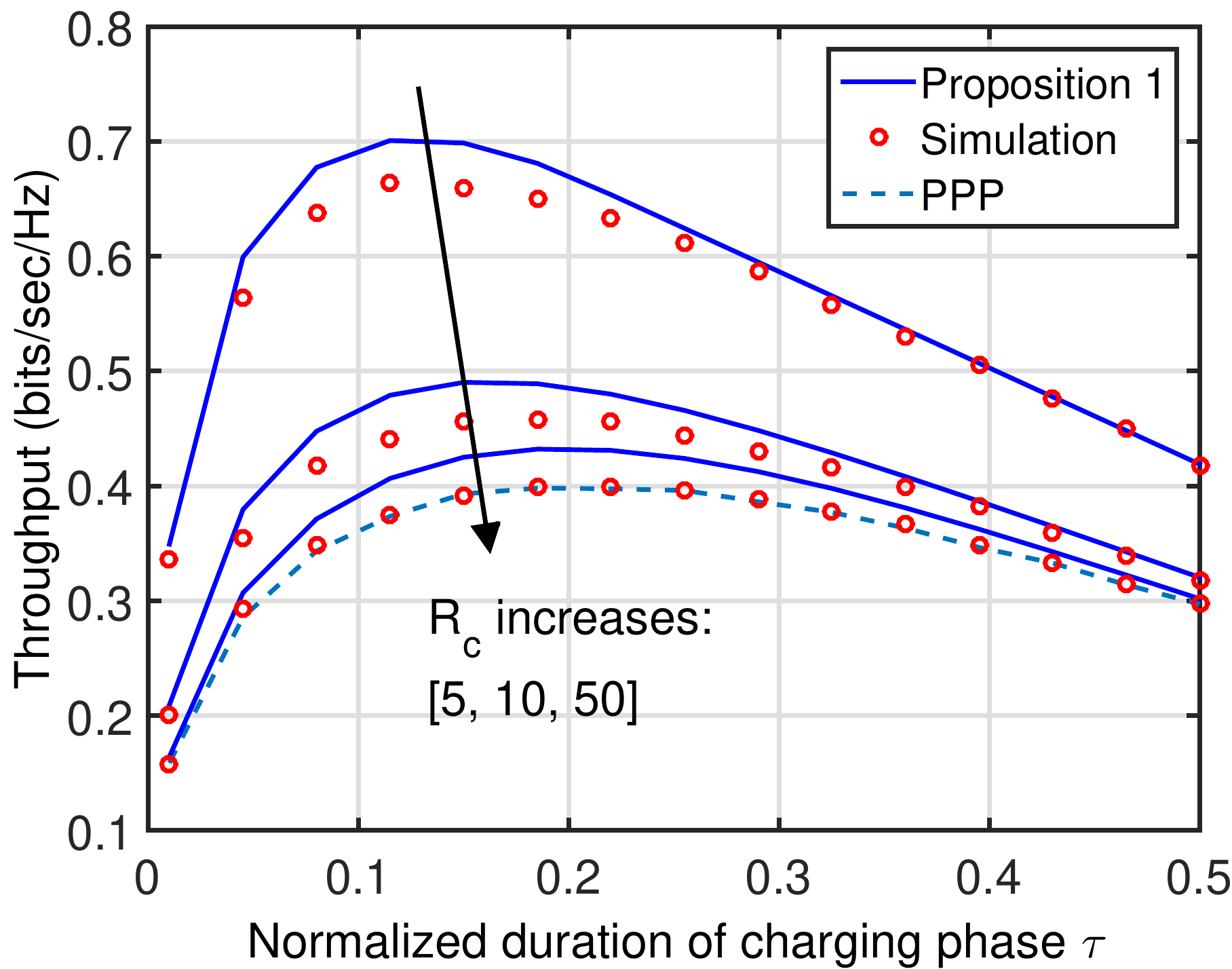}
\caption{The average downlink achievable throughput of a typical IoT device when the representative cluster has a GW deployed at its center for Mat{\'e}rn cluster process.}
\label{fig:6}
    \end{minipage}
\end{figure}
\begin{figure}[!t]
    \centering
    \begin{minipage}{.41\textwidth}
        \centering
\includegraphics[width=1\columnwidth]{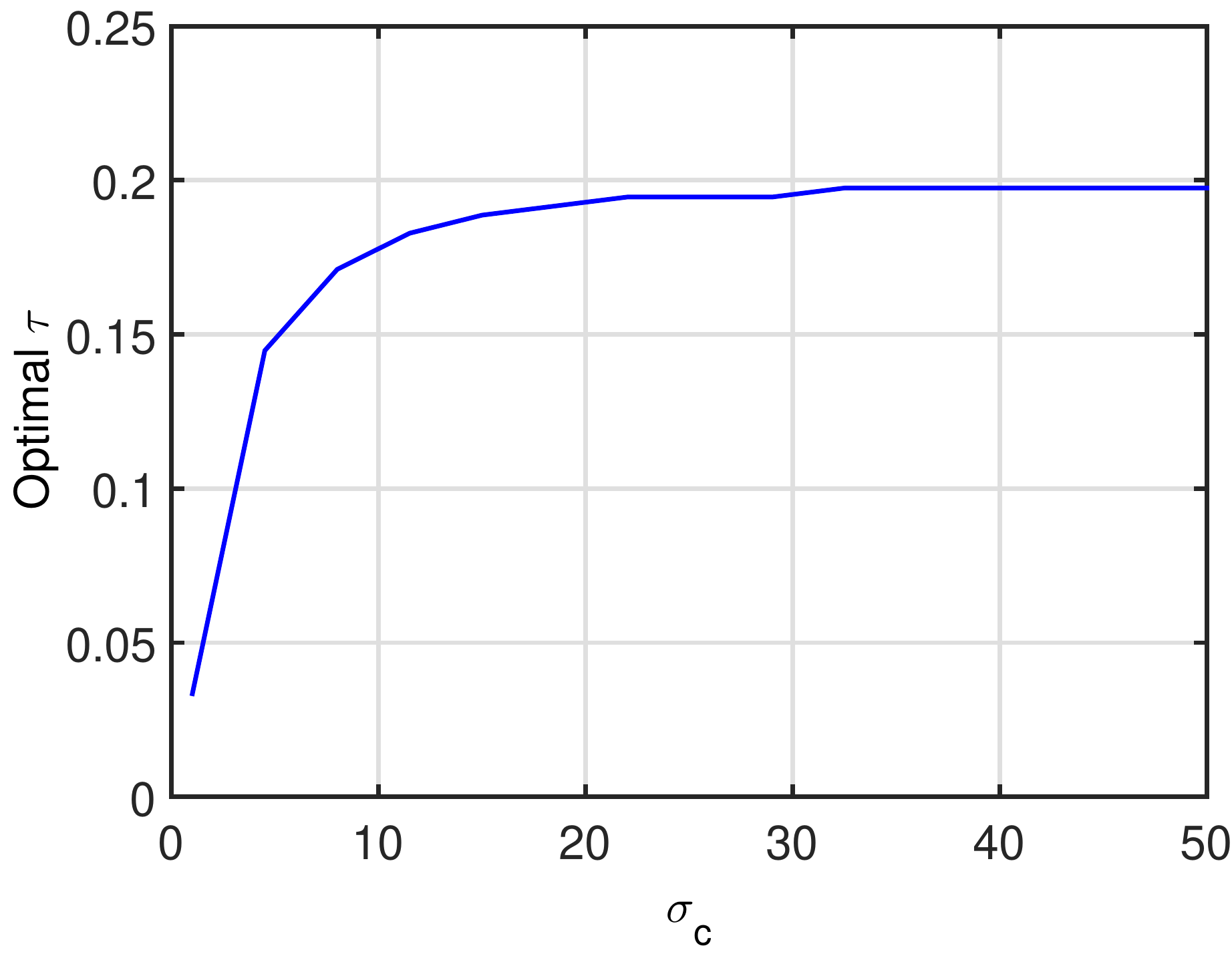}
\caption{The value of optimal $\tau$ for different values of $\sigma_c$ when the representative cluster has a GW deployed at its center for Thomas cluster process.}
\label{fig:7}
    \end{minipage}%
    \hfill
    \begin{minipage}{0.41\textwidth}
        \centering
\includegraphics[width=1\columnwidth]{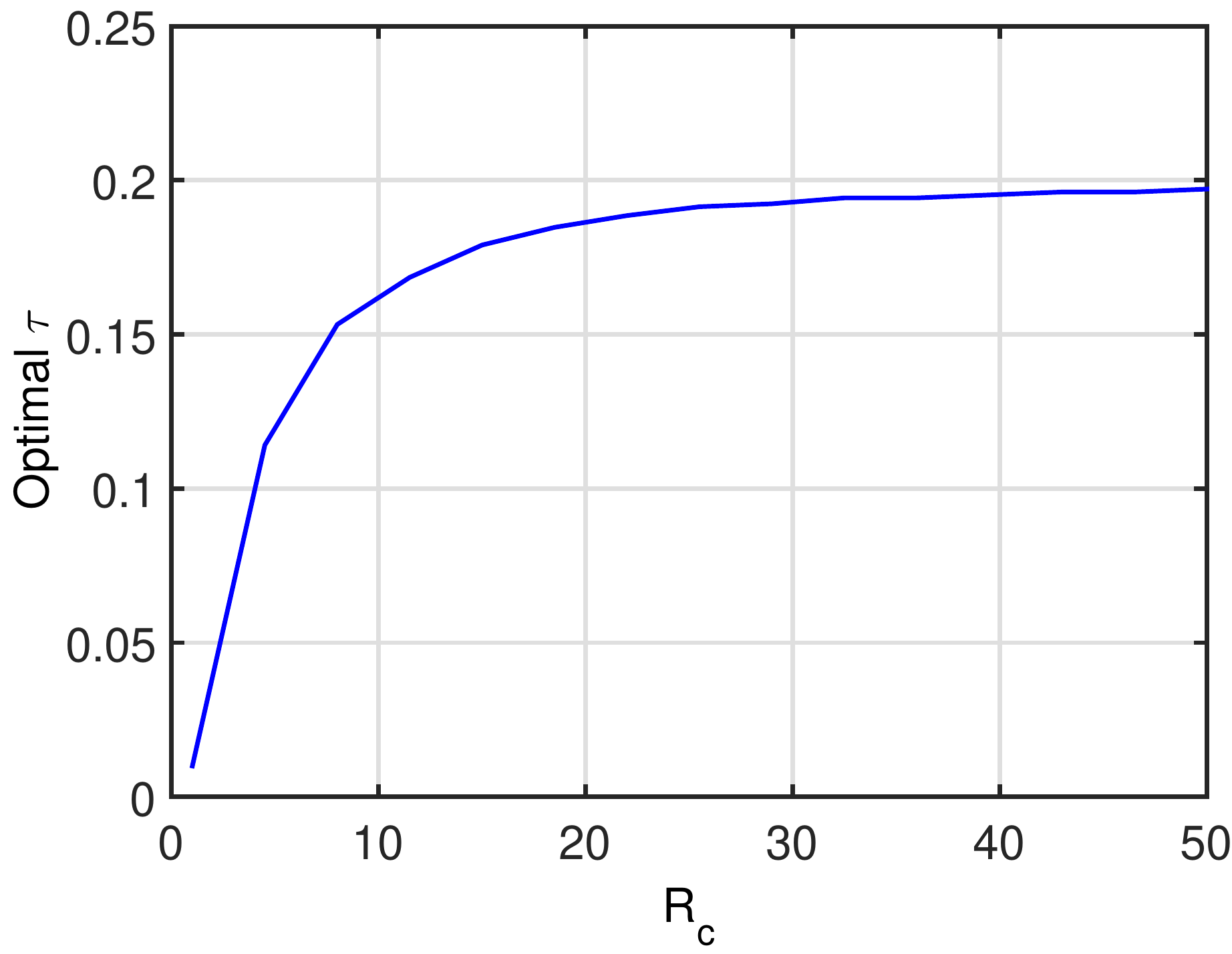}
\caption{The value of optimal $\tau$ for different values of ${\rm R}_c$ when the representative cluster has a GW deployed at its center for Mat{\'e}rn cluster process.}
\label{fig:8}
    \end{minipage}
\end{figure}

In Figs.~\ref{fig:5} and~\ref{fig:6} we plot the average throughput provided in Proposition~\ref{prop:1}. We observe the existence of an optimal value of $\tau$ that maximizes the throughput, as already discussed in Remark~\ref{rem:9}. The optimal values of $\tau$ for both Thomas and Mat{\'e}rn cluster process are plotted in Figs.~\ref{fig:7} and~\ref{fig:8}. We note that this optimal value converges to a certain fixed value as the cluster size increases, which is the optimal value of $\tau$ when the locations of the IoT devices and GWs are modeled by two independent PPPs.

\section{Conclusion} 
In this paper, we provided the performance analysis of a generalized system setup of RF-powered IoT that captures the coupling between the locations of the IoT devices and the locations of the RF sources. In particular, we studied a system setup constructed of two networks: (i) the IoT, where the locations of the IoT devices are modeled by PCP and (ii) the wireless network that powers the IoT devices, where a fraction of the total GWs are deployed at the cluster centers and the rest are randomly located in the 2-D plane. The system setup considered in this paper can be tuned to capture any level of coupling between the locations of the IoT devices and the locations of the RF sources (the GWs). For this setup, we derived the energy coverage and the joint coverage probability of the IoT device in the downlink. We proposed two different approaches to handle the derivation challenges that result from modeling the locations of the IoT devices using PCP. 

Multiple system insights were drawn using the expressions derived in this paper.
For instance, from the energy coverage perspective, the performance of the system setup considered in this paper (IoT devices clustered around the GWs) is lower bounded by the performance of the setup in which the locations of the IoT devices are completely independent from the locations of the BSs. This observation implies that, in addition to capacity enhancement and patching coverage dead-zones, deployment of GWs at high-density areas (i.e clusters) significantly affects the energy harvesting performance of RF-powered IoT devices as well. Our results also showed that the optimal slot partitioning is fairly sensitive to the cluster size and the fraction of total GWs that are deployed at the cluster centers. 

This work has many possible extensions. For instance, we focused in this paper only on the downlink performance of the clustered RF-powered IoT. One possible extension would be to consider the joint uplink/downlink coverage probability of this system setup. In addition, another possible extension is to consider battery-equipped IoT devices with finite battery sizes. In that case, the dynamics and steady state distribution of the battery levels would explicitly appear in the analysis.
Another possible future work is incorporating the assumption of having energy harvesting circuitries with non-linear efficiency.
\appendix

\subsection{Proof of Lemma~\ref{lem:E_cov_tier1_approx1}} \label{app:E_cov_tier1_approx1}
The energy coverage probability, conditioned on $\Phi$ and the association of the typical IoT device with $\Phi_1$, can be expressed as follows
\begin{align}\label{E_cov_tier1}
&E_{{\rm cov} \g \Phi}^{(1)} \nonumber = \P\left(E_{\rm H} \geq E_{\rm rec}\g {\rm index}=1, \Phi\right)\nonumber\\&= \P\left( \eta \tau T \sum_{\substack{\nbx \in \Phi}}{P_{\nrmt} g_{\nbx} \lVert \nbx \rVert ^{-\alpha}} \geq E_{\rm rec} \Bg {\rm index}=1, \Phi\right)\nonumber\\
 & \overset{(a)}{\approx} \P\Bigg(\E\left[\sum_{\substack{\nbx \in \Phi \setminus \nbx^* }}{ g_{\nbx} \lVert \nbx \rVert ^{-\alpha}} \Bg {\rm index}=1, \lVert \nbx^* \rVert \right] \nonumber\\&+g_{\nbx^*} \lVert \nbx^* \rVert^{-\alpha}\geq C(\tau) \Bg {\rm index}=1, \Phi\Bigg)\nonumber\\
& \b \P\Bigg( \E\left[\sum_{\substack{\nbx_{1} \in \Phi_{\rm 1} \setminus \nbx_1^* }}{ g_{\nbx_{1}} \lVert \nbx_{1} \rVert ^{-\alpha}} + g_{\nbx_{0}} \lVert \nbx_{0} \rVert ^{-\alpha} \Bg {\rm index}=1, w_{1} \right]\nonumber\\& +g_{\nbx_{1}^*} w_1^{-\alpha} \geq C(\tau) \Bg w_{1}\Bigg)\nonumber\\
 & = \P\left( g_{\nbx_{1}^*} w_1^{-\alpha} + \Psi(w_1) \geq C(\tau) \g w_{1}\right)\nonumber\\& \overset{(c)}{=}{\rm e}^{-\left[w_{1}^{\alpha} \left(C(\tau) - \Psi(w_1)\right)\right]^{+}},
\end{align}
where in step (a), under Approximation $1$, the energy harvested at the typical device is approximated by the sum of energy harvested from its serving GW $\nbx^*$ and the conditional mean of the energy harvested from the other GWs, and $C(\tau) = \frac{E_{\rm rec}}{\eta \tau T P_{\nrmt}}$. Step (b) follows from the conditioning on the association with $\Phi_1$ and (c) follows from the Rayleigh fading assumption, i.e., $g_{\nbx_1^*} \sim {\rm exp}(1)$.

The conditional mean of the energy harvested from all GWs except the serving one is derived as follows $\Psi(w_1)= $
\begin{align}\label{E_mean_tier1}
&\E\left[\sum_{\substack{\nbx_{1} \in \Phi_{\rm 1} \setminus \nbx_1^* }}{ g_{\nbx_{1}} \lVert \nbx_{1} \rVert ^{-\alpha}} + g_{\nbx_{0}} \lVert \nbx_{0} \rVert ^{-\alpha} \Bg {\rm index}=1, w_{1} \right]\nonumber\\
 &\a \E_{R_{0}}\left[\lVert \nbx_0 \rVert ^{-\alpha} \g R_{0} > w_{1}\right] + \E_{\Phi_{1}}\left[\sum_{\substack{\nbx_1 \in \Phi_1 \setminus  \nbx_1^*}}{\lVert \nbx_1 \rVert ^ {-\alpha}} \Bg w_1\right]\nonumber\\
&\b \int_{r_0 > w_1}^{\infty} r_{0}^{-\alpha}\frac{f_{R_{0}}(r_0)}{\bar{F}_{R_{0}}(w_1)} \nrmd r_{0} + 2 \pi \lambda \int_{w_1}^{\infty} r^{-\alpha} r \nrmd r  \nonumber\\&= \int_{r_0 > w_1}^{\infty} r_{0}^{-\alpha}\frac{f_{R_{0}}(r_0)}{\bar{F}_{R_{0}}(w_1)} \nrmd r_{0} + \frac{2 \pi \lambda}{\alpha - 2} w_1^{2 - \alpha},
\end{align}
where the expectation is performed over $\Phi_{\rm 1} \setminus \nbx_1^*$ and $\nbx_0$ while deriving $\Psi(w_1)$, (a) follows from the fact that the channel gains are assumed to be Rayleigh distributed along with conditioning on the event that the typical device is associated with $\Phi_1$ which implies that $R_0 >  w_1$, and (b) follows from Campbell's Theorem~\cite{haenggi2012stochastic} with conversion from Cartesian to polar coordinates. Finally, the unconditional energy coverage probability, given the association with $\Phi_1$, i.e., $E_{\rm cov}^{(1)}$, is obtained by taking the expectation of (\ref{E_cov_tier1}) with respect to the serving distance $W_{1}$. This completes the proof.\hfill 
\IEEEQED

\subsection{Proof of Lemma~\ref{lem:E_cov_tier0_approx1}} \label{app:E_cov_tier0_approx1}
Given that the typical device is associated with $\Phi_0$, i.e., the GW located at its cluster center, the energy coverage probability conditioned on $\Phi$ is obtained as follows
\begin{align}\label{E_cov_tier0}
&E_{{\rm cov} \g \Phi}^{(0)} \nonumber = \P\left(E_{\rm H} \geq E_{\rm rec}\g {\rm index}=0, \Phi\right)\nonumber\\
 & \overset{(a)}{\approx} \P\Bigg( g_{\nbx^*} \lVert \nbx^* \rVert^{-\alpha} + \E\left[\sum_{\substack{\nbx \in \Phi \setminus \nbx* }}{ g_{\nbx} \lVert \nbx \rVert ^{-\alpha}} \Bg {\rm index}=0, \lVert \nbx^* \rVert \right] \nonumber\\&\geq C(\tau) \Bg {\rm index}=0, \Phi\Bigg)\nonumber\\
&= \P\left( g_{\nbx_{0}} w_0^{-\alpha} + \E\left[\sum_{\substack{\nbx_{1} \in \Phi_{\rm 1}}}{ g_{\nbx_{1}} \lVert \nbx_{1} \rVert ^{-\alpha}} \Bg w_{0} \right] \geq C(\tau) \Bg w_{0}\right)\nonumber\\
 & = \P\left( g_{\nbx_{0}} w_0^{-\alpha} + \theta(w_0) \geq C(\tau) \g w_{0}\right) \overset{(b)}{=}{\rm e}^{-\left[w_{0}^{\alpha} \left(C(\tau) - \theta(w_0)\right)\right]^{+}},
\end{align}
where (a) follows from approximating the energy harvested by the typical IoT device under Approximation $1$ and (b) follows from the fact that $g_{\nbx_0} \sim {\rm exp}(1)$. In addition, the conditional mean of the harvested energy from all GWs except the serving one can be derived as follows
\begin{align}\label{E_mean_tier0}
\theta(w_0) &= \E\left[\sum_{\substack{\nbx_{1} \in \Phi_{\rm 1}}}{ g_{\nbx_{1}} \lVert \nbx_{1} \rVert ^{-\alpha}} \Bg w_{0} \right] \nonumber\\&\a \E_{\Phi_{1}}\left[\sum_{\substack{\nbx_1 \in \Phi_1}}{\lVert \nbx_1 \rVert ^ {-\alpha}} \Bg w_0\right] \nonumber\\&\b  2 \pi \lambda \int_{w_0}^{\infty} r^{-\alpha} r \nrmd r  = \frac{2 \pi \lambda}{\alpha - 2} w_0^{2 - \alpha},
\end{align}
where (a) follows from the Rayleigh fading assumption and (b) follows from Campbell's Theorem~\cite{haenggi2012stochastic} for sum over PPP with the transformation to polar coordinates. Substituting $\theta(w_0)$ from (\ref{E_mean_tier0}) into (\ref{E_cov_tier0}), we obtain
\begin{align}\label{condition_E_cov_tier0}
E_{{\rm cov} \g \Phi}^{(0)} = 
          \begin{cases}
           \nrme^{-\left(C(\tau) w_0^{\alpha} - \frac{2 \pi \lambda}{\alpha - 2} w_0^2\right)},\;{\rm if} \; w_0 \geq A \\
            1,\;{\rm if} \; w_0 < A
          \end{cases}
\end{align}
where $A = \left(\frac{2 \pi \lambda}{C(\tau) \left(\alpha - 2\right)}\right)^{\frac{1}{\alpha - 2}}$. The expression of the unconditional energy coverage probability in (\ref{eq:E_cov_tier0_approx1}) follows from taking the expectation over the serving distance $W_0$ along with applying the condition in (\ref{condition_E_cov_tier0}). This completes the proof.\hfill 
\IEEEQED

\subsection{Proof of Lemma~\ref{lem:E_cov_tier1_approx2}}
\label{app:E_cov_tier1_approx2}
Given the association of the typical IoT device with $\Phi_1$, the energy coverage probability conditioned on $\Phi$ can be obtained as follows
\begin{align}\label{E_cov_tier1_approx2}
&E_{{\rm cov} \g \Phi}^{(1)} \nonumber = \P\left(E_{\rm H} \geq E_{\rm rec}\g {\rm index}=1, \Phi\right)\\&\nonumber= \P\left( \eta \tau T \sum_{\substack{\nbx \in \Phi}}{P_{\nrmt} g_{\nbx} \lVert \nbx \rVert ^{-\alpha}} \geq E_{\rm rec} \Bg {\rm index}=1, \Phi\right)\\
\nonumber & \overset{(a)}{\approx} \P\Big( g_{\nbx_1^*} \lVert \nbx_1^* \rVert^{-\alpha} + g_{\nbx_0} \lVert \nbx_0 \rVert^{-\alpha} + \Psi(w_1)\\&\nonumber \geq C(\tau) \g {\rm index}=1, \Phi\Big)\\
\nonumber & \b \P\Big( g_{\nbx_{1}^*} w_1^{-\alpha} + g_{\nbx_{0}} r_0^{-\alpha} \\&\nonumber \geq C(\tau) - \Psi(w_1) \g {\rm index}=1, w_{1}, r_0 \Big)\\
&\overset{(c)}{=}\frac{w_1^{-\alpha} \nrme^{- w_1^{\alpha} \left[C(\tau) - \Psi (w_1)\right]^+} - r_0^{-\alpha} \nrme^{- r_0^{\alpha} \left[C(\tau) - \Psi (w_1)\right]^+}}{w_1^{-\alpha} - r_0^{-\alpha}},
\end{align}
where in step (a), under Approximation $2$, the energy harvested at the typical device is approximated by the sum of energy harvested from $\nbx_1^*$, $\nbx_0$ and the conditional mean of the energy harvested from the other GWs, and $\Psi(w_1) = \E\left[\sum_{\substack{\nbx \in \Phi \setminus \nbx_1^*, \nbx_0 }}{ g_{\nbx} \lVert \nbx \rVert ^{-\alpha}} \g {\rm index}=1, \lVert \nbx_1^* \rVert, \lVert \nbx_0 \rVert \right]$. Step (b) follows from the conditioning on the association with $\Phi_1$ and (c) is due to hypo-exponential distribution of $g_{\nbx_{1}^*} w_1^{-\alpha} + g_{\nbx_{0}} r_0^{-\alpha}$ (sum of two independent exponential random variables, $g_{\nbx_{1}^*}$ and $g_{\nbx_{0}}$, with means $w_1^{\alpha}$ and $r_0^{\alpha}$, respectively). By applying similar analysis as in (\ref{E_mean_tier0}), the value of $\Psi(w_1)$ can be obtained as
\begin{align}\label{epsi_second_approximation}
\Psi(w_1) = \dfrac{2 \pi \lambda}{\alpha - 2} w_1^{2 - \alpha}.
\end{align}

From (\ref{E_cov_tier1_approx2}), the unconditional energy coverage probability, given the association with $\Phi_1$, can be expressed as
\begin{align}
E_{\rm cov}^{(1)} \nonumber &= \E\Bigg[\frac{w_1^{-\alpha} \nrme^{- w_1^{\alpha} \left[C(\tau) - \Psi (w_1)\right]^+}}{w_1^{-\alpha} - r_0^{-\alpha}}\\&\nonumber-\frac{r_0^{-\alpha} \nrme^{- r_0^{\alpha} \left[C(\tau) - \Psi (w_1)\right]^+}}{w_1^{-\alpha} - r_0^{-\alpha}}\Bg {\rm index}=1\Bigg]\\
\nonumber &\a \E_{W_1}\E_{R_0}\Bigg[\frac{w_1^{-\alpha} \nrme^{- w_1^{\alpha} \left[C(\tau) - \Psi (w_1)\right]^+}}{w_1^{-\alpha} - r_0^{-\alpha}}\\&\nonumber-\frac{r_0^{-\alpha} \nrme^{- r_0^{\alpha} \left[C(\tau) - \Psi (w_1)\right]^+}}{w_1^{-\alpha} - r_0^{-\alpha}}\Bg R_0 > w_1\Bigg]\\
\nonumber &= \E_{W_1}\Bigg[\int_{w1}^{\infty}\Bigg(\frac{w_1^{-\alpha} \nrme^{- w_1^{\alpha} \left[C(\tau) - \Psi (w_1)\right]^+}}{w_1^{-\alpha} - r_0^{-\alpha}}\\&\nonumber-\frac{r_0^{-\alpha} \nrme^{- r_0^{\alpha} \left[C(\tau) - \Psi (w_1)\right]^+}}{w_1^{-\alpha} - r_0^{-\alpha}}\Bigg) \times\frac{f_{R_0}(r_0)}{\bar{F}_{R_0}(w_1)} \nrmd r_0 \Bigg]\\
\nonumber &\b \frac{1}{A_1}\int_{0}^{A}\int_{w_1}^{\infty}{f_{R_0}(r_0) f_{R_1}(w_1)\nrmd r_0 \nrmd w_1} \\
&+ \frac{1}{A_1}\int_{A}^{\infty}\int_{w_1}^{\infty}\Bigg(\frac{w_1^{-\alpha} \nrme^{- w_1^{\alpha} \left[C(\tau) - \Psi (w_1)\right]}}{w_1^{-\alpha} - r_0^{-\alpha}}\\&\nonumber-\frac{r_0^{-\alpha} \nrme^{- r_0^{\alpha} \left[C(\tau) - \Psi (w_1)\right]}}{w_1^{-\alpha} - r_0^{-\alpha}}\Bigg)f_{R_0}(r_0)f_{R_1}(w_1) \nrmd r_0 \nrmd w_1,
\end{align}
where (a) follows by distributing the expectation over the random quantities while taking into account the conditioning on association with $\Phi_1$ which implies that $R_0 > w_0$ and (b) results from substituting the value of $\Psi(w_1)$ from (\ref{epsi_second_approximation}) along with taking the expectation over the serving distance $W_1$. The final result is obtained by observing that the first term in step (b) is equal to $F_{W_1}(A)$.\hfill 
\IEEEQED
\subsection{Proof of Lemma~\ref{lem:joint_cov_tier1}}
\label{app:joint_cov_tier1}
Given that the typical device is associated with $\Phi_1$, the joint coverage probability is obtained as follows
\begin{align}\label{joint_cov_tier1}
P_{\rm cov}^{(1)} \nonumber &= \E\left[\mathbbm{1}\left(SINR \geq \T \right)\mathbbm{1}\left(E_{\rm H} \geq E_{\rm rec}\right)\g {\rm index} = 1\right]\\ 
\nonumber &\a \E_{\Phi}\Bigg[\E_{h}\left[\mathbbm{1}\left(SINR \geq \T \right) \g {\rm index}=1, \Phi\right] \times\nonumber\\&\E_{g}\left[ \mathbbm{1}\left(E_{\rm H} \geq E_{\rm rec}\right) \g {\rm index}=1, \Phi\right]\Bigg]\nonumber\\
&= \E_{\Phi}\Bigg[\P\left(SINR \geq \T \g {\rm index}=1, \Phi \right) \times\nonumber\\& \P\left(E_{\rm H} \geq E_{\rm rec}\g {\rm index}=1, \Phi\right)\Bigg],
\end{align}
where (a) follows from the fact that the energy and SINR coverage events are (conditionally) independent conditioned on $\Phi$. The SINR coverage probability conditioned on $\Phi$ and the association with $\Phi_1$ is derived as follows
\begin{align}\label{I_cov_tier1}
&S_{{\rm cov}\g \Phi}^{(1)} \nonumber = \P\left(SINR \geq \T \g {\rm index}=1, \Phi \right) \\&\nonumber=  \P\left(\frac{P_{\nrmt} h_{\nbx_{1}^*} w_1^{-\alpha}}{I_{1} + \N} \geq \T \bg {\rm index}=1, \Phi \right)\\
\nonumber & \a \E_{h_{\nbx_{0}},h_{\nbx_1}}\left[{\rm exp}\left(-\frac{\T w_1^{\alpha}\left(I_{1} + \N \right)}{P_{\nrmt}}\right)\right]\\&\nonumber \b \nrme^{-\frac{\T \N w_1^{\alpha}}{P_{\nrmt}}} \E_{h_{\nbx_0}}\left[\nrme^{- \T w_1^{\alpha} \lVert \nbx_{0} \rVert ^{-\alpha} h_{\nbx_{0}}}\right] \times\\&\nonumber\prod_{\nbx_1 \in \Phi_1 \setminus \nbx_1^* }{\E_{h_{\nbx_1}}\left[\nrme^{- \T w_1^{\alpha} \lVert \nbx_{1} \rVert ^{-\alpha} h_{\nbx_{1}}}\right]}\\
& \c \nrme^{-\frac{\T \N w_1^{\alpha}}{P_{\nrmt}}} \frac{1}{1 + \T w_1^{\alpha} \lVert \nbx_0 \rVert^{-\alpha}} \prod_{\nbx_1 \in \Phi_1 \setminus \nbx_1^* }{\frac{1}{1 + \T w_1^{\alpha} \lVert \nbx_1 \rVert^{-\alpha}}},
\end{align}
where $I_{i} = \sum_{\substack{\nbx \in \Phi \setminus} \nbx^*_i}{P_{\nrmt} h_{\nbx} \lVert \nbx \rVert ^{-\alpha}}$, (a) follows from the fact that $h_{\nbx_1^*}\sim {\rm exp}(1)$, (b) follows from the independence of the channel power gains $h_{\nbx_0}$ and $\{h_{\nbx_1}\}$, and (c) follows from the assumption that the channel gains are Rayleigh distributed. Therefore, from (\ref{eq:E_cov_tier1_approx1_phi}) and (\ref{I_cov_tier1}), the joint coverage probability conditioned on the association with $\Phi_1$, defined in (\ref{joint_cov_tier1}), can be expressed as
\begin{align}\label{Eq: usedin_bounds_proof}
P_{\rm cov}^{(1)} \nonumber &= \E_{\Phi}\left[E_{{\rm cov} \g \Phi}^{(1)} S_{{\rm cov }\g \Phi}^{(1)} \bg {\rm index}=1 \right]\\ 
\nonumber &= \E_{\Phi}\Bigg[ \nrme^{-\frac{\T \N w_1^{\alpha}}{P_{\nrmt}}} \frac{1}{1 + \T w_1^{\alpha} \lVert \nbx_0 \rVert^{-\alpha}} \times\\&\nonumber\prod_{\nbx_1 \in \Phi_1 \setminus \nbx_1^* }{\frac{1}{1 + \T w_1^{\alpha} \lVert \nbx_1 \rVert^{-\alpha}}} {\rm e}^{-\left[w_{1}^{\alpha} \left(C(\tau) - \Psi(w_1)\right)\right]^{+}} \Bg {\rm index}=1\Bigg]\\
\nonumber &\a \E_{w_1}\Bigg[\nrme ^{-\left( \frac{\T \N w_1^{\alpha}}{P_{\nrmt}} + \left[w_{1}^{\alpha} \left(C(\tau) - \Psi(w_1)\right)\right]^{+}\right)} \times\\&\nonumber\E_{R_0}\left[\frac{1}{1 + \T w_1^{\alpha} r_0^{-\alpha}} \bg R_0 > w_1\right] \times\\&\nonumber \E_{\Phi_1 \setminus \nbx_1^*}\Bigg[\prod_{\nbx_1 \in \Phi_1 \setminus \nbx_1^* }\frac{1}{1 + \T w_1^{\alpha} \lVert \nbx_1 \rVert^{-\alpha}}\Bigg] \Bigg]\\
 \nonumber&\b \E_{w_1}\Bigg[\nrme ^{-\left( \frac{\T \N w_1^{\alpha}}{P_{\nrmt}} + \left[w_{1}^{\alpha} \left(C(\tau) - \Psi(w_1)\right)\right]^{+} + 2 \pi \lambda w_1^2 \rho(\T,\alpha)\right)} \times\\&\nonumber \int_{r_0 > w_1}^{\infty} \frac{1}{1 + \T w_1^{\alpha} r_0^{-\alpha}}\frac{f_{R_{0}}(r_0)}{\bar{F}_{R_{0}}(w_1)} \nrmd r_{0}\Bigg]\\
\nonumber &\c \frac{1}{A_1}\int_{w_1 = 0}^{\infty} \nrme ^{-\left( \frac{\T \N w_1^{\alpha}}{P_{\nrmt}} + \left[w_{1}^{\alpha} \left(C(\tau) - \Psi(w_1)\right)\right]^{+} + 2 \pi \lambda w_1^2 \rho(\T,\alpha)\right)}\nonumber\\& \int_{r_0 > w_1}^{\infty} \frac{1}{1 + \T w_1^{\alpha} r_0^{-\alpha}}f_{R_{0}}(r_0)\nrmd r_{0} \nrmd w_1,
\end{align}
where (a) follows by distributing the expectation over the point process $\Phi_1 \setminus \nbx_1^*$ and the rest of random quantities, (b) follows from the PGFL of the PPP~\cite{haenggi2012stochastic} where $\rho(\T,\alpha) = \frac{\T ^{\frac{2}{\alpha}}}{2} \int_{\T^{\frac{-2}{\alpha}}}^{\infty}{\frac{1}{1 + u^{\frac{\alpha}{2}}} \nrmd u}$ and (c) follows from (\ref{W1_pdf}). This completes the proof. 
\hfill 
\IEEEQED

\subsection{Proof of Lemma~\ref{lem:joint_cov_tier0}}
\label{app:joint_cov_tier0}
 The SINR coverage probability conditioned on $\Phi$ and the association with $\Phi_0$ can be obtained as follows
\begin{align}\label{I_cov_tier0}
S_{{\rm cov}\g \Phi}^{(0)} \nonumber &= \P\left(SINR \geq \T \g {\rm index}=0, \Phi \right) \\&\nonumber=  \P\left(\frac{P_{\nrmt} h_{\nbx_{0}} w_0^{-\alpha}}{I_{0} + \N} \geq \T \bg {\rm index}=0, \Phi \right)\\
\nonumber & \a \E_{h_{\nbx_1}}\left[{\rm exp}\left(-\frac{\T w_0^{\alpha}\left(I_{0} + \N \right)}{P_{\nrmt}}\right) \Bg {\rm index}=0, \Phi\right]\\&\nonumber\b \nrme^{-\frac{\T \N w_0^{\alpha}}{P_{\nrmt}}} \prod_{\nbx_1 \in \Phi_1}{\E_{h_{\nbx_1}}\left[\nrme^{- \T w_0^{\alpha} \lVert \nbx_{1} \rVert ^{-\alpha} h_{\nbx_{1}}}\right]}\\
& \c \nrme^{-\frac{\T \N w_0^{\alpha}}{P_{\nrmt}}} \prod_{\nbx_1 \in \Phi_1 }{\frac{1}{1 + \T w_0^{\alpha} \lVert \nbx_1 \rVert^{-\alpha}}}
\end{align}
where (a) follows from the fact that $h_{\nbx_0}\sim {\rm exp}(1)$, (b) follows from the independence of the channel power gains between the interfering GWs and the typical IoT device and (c) follows from $h_{\nbx_1}\sim {\rm exp}(1)$. Therefore, from (\ref{eq:E_cov_tier0_approx1_Phi}) and (\ref{I_cov_tier0}), the joint coverage probability conditioned on the association with $\Phi_0$ can be expressed as
\begin{align}\label{joint_cov_tier0}
P_{\rm cov}^{(0)} \nonumber &= \E_{\Phi}\left[E_{{\rm cov} \g \Phi}^{(0)} S_{{\rm cov }\g \Phi}^{(0)} \g {\rm index}=0\right]\\ 
\nonumber &= \E_{\Phi}\Bigg[ \nrme^{-\frac{\T \N w_0^{\alpha}}{P_{\nrmt}}}\times\\&\nonumber \prod_{\nbx_1 \in \Phi_1}{\frac{1}{1 + \T w_0^{\alpha} \lVert \nbx_1 \rVert^{-\alpha}}} {\rm e}^{-\left[w_{0}^{\alpha} \left(C(\tau) - \theta(w_0)\right)\right]^{+}}\Bg {\rm index}=0\Bigg]\\
\nonumber &\a \E_{w_0}\Bigg[\nrme ^{-\left( \frac{\T \N w_0^{\alpha}}{P_{\nrmt}} + \left[w_{0}^{\alpha} \left(C(\tau) - \theta(w_0)\right)\right]^{+}\right)} \\&\nonumber\E_{\Phi_1}\left[\prod_{\nbx_1 \in \Phi_1}\frac{1}{1 + \T w_0^{\alpha} \lVert \nbx_1 \rVert^{-\alpha}} \Bg {\rm index}=0 \right] \Bigg]\\
\nonumber &\b \E_{w_0}\left[\nrme ^{-\left( \frac{\T \N w_0^{\alpha}}{P_{\nrmt}} + \left[w_{0}^{\alpha} \left(C(\tau) - \theta(w_0)\right)\right]^{+} + 2 \pi \lambda w_0^2 \rho(\T,\alpha)\right)}\right]\\
&\c \int_{0}^{A} \nrme ^{-\left( \frac{\T \N w_0^{\alpha}}{P_{\nrmt}} + 2 \pi \lambda w_0^2 \rho(\T,\alpha)\right)} f_{W_{0}}(w_0)\nrmd w_0\\&\nonumber + \int_{A}^{\infty} {\nrme^{-\left(\left[\frac{\T \N}{P_{\nrmt}} + C(\tau)\right]w_0^{\alpha} +  \left[\rho(\T,\alpha) - \frac{1}{\alpha - 2}\right] 2 \pi \lambda w_0^2\right)} f_{W_{0}}(w_0)} \nrmd w_0, 
\end{align}
where (a) following by distributing the expectation over different random quantities, (b) follows from the PGFL of the PPP~\cite{haenggi2012stochastic} and (c) follows from applying the condition in (\ref{condition_E_cov_tier0}). This completes the proof.
\bibliographystyle{IEEEtran}
\bibliography{TGCN_submitted}
\end{document}